\documentclass[a4paper,11pt]{article}

\usepackage{jheppub}
\usepackage[T1]{fontenc}
\usepackage{graphicx}
\usepackage{amsmath}
\usepackage{amsmath}
\usepackage{xspace}
\usepackage{subfig}
\usepackage{color}
\usepackage{xcolor}
\usepackage[utf8]{inputenc}
\usepackage{commath}
 \usepackage{cancel}
\usepackage{slashed}
\usepackage{fancyvrb}
\usepackage{multirow}

\title{Probing Heavy Neutrinos at the LHC from Fat-jet using Machine Learning}

\author[a]{Wei Liu,}
\author[b]{Jing Li,}
\author[a]{Zixiang Chen,}
\author[b,1]{Hao Sun}
\footnotetext[1]{Corresponding author.}

\affiliation[a]{Department of Applied Physics and MIIT Key Laboratory of Semiconductor Microstructure and Quantum Sensing, Nanjing University of Science and Technology, Nanjing 210094, China}
\affiliation[b]{Institute of Theoretical Physics, School of Physics, Dalian University of Technology, No.2 Linggong Road, Dalian, Liaoning, 116024, P.R.China }

\emailAdd{wei.liu@njust.edu.cn}
\emailAdd{haosun@dlut.edu.cn}

\begin{abstract}{
We explore the potential to use machine learning methods to search for heavy neutrinos, from their hadronic final states including a fat-jet signal, via the processes $pp \rightarrow W^{\pm *}\rightarrow \mu^{\pm} N \rightarrow \mu^{\pm} \mu^{\mp} W^{\pm} \rightarrow \mu^{\pm} \mu^{\mp} J$ at hadron colliders. We use either the Gradient Boosted Decision Tree or Multi-Layer Perceptron methods to analyse the observables incorporating the jet substructure information, which is performed at hadron colliders with $\sqrt{s}=$ 13, 27, 100 TeV. It is found that, among the observables, the invariant masses of variable system and the observables from the leptons are the most powerful ones to distinguish the signal from the background. With the help of machine learning techniques, the limits on the active-sterile mixing have been improved by about one magnitude comparing to the cut-based analyses, with $V_{\mu N}^2 \lesssim 10^{-4}$ for the heavy neutrinos with masses, 100 GeV$~<m_N<~$1 TeV.}
\end{abstract}

\begin{document}
\maketitle
\flushbottom
\setcounter{footnote}{0}

\section{Introduction}
The observation of tiny neutrino masses is one of the most direct evidence for the existence of the new physics beyond the Standard Model~(SM). It can be explained by the canonical type-I seesaw, but the Dirac Yukawa couplings required by the eV scale active neutrino masses are not natural, $Y_D \sim 10^{-6}$, if the right-handed~(heavy) neutrinos are in electroweak~(EW) scale~\cite{Mohapatra:1979ia,Yanagida:1979as,Gell-Mann:1979vob,Deppisch:2015qwa}. This results in the suppression on the active-sterile mixing~($V_{lN}$), making the heavy neutrinos long-lived, which received a lot of attention~\cite{Deppisch:2018eth,Amrith:2018yfb,Deppisch:2019kvs,Liu:2022kid,Liu:2022ugx}.
Such suppression can be lifted away, if the inverse seesaw is considered where the smallness of the neutrino masses can be absorbed by a naturally small lepton number violating parameter, while additional SM singlet fermions $S$ are also introduced~\cite{Mohapatra:1986aw,Mohapatra:1986bd}. Thus, the Yukawa couplings can be of order one, and the mixing between the active and heavy neutrinos $V_{lN}$ can be sizeable, with $V_{lN}^2 \lesssim 10^{-3}$ from electroweak precision data~(EWPD)~\cite{delAguila:2008pw,Akhmedov:2013hec,Antusch:2014woa,Blennow:2016jkn},
leading to very rich phenomenology, especially at colliders~\cite{Abdullahi:2022jlv,Liu:2021akf,Balaji:2020oig,Liu:2023nxi,Zhang:2023nxy,Barducci:2022gdv,Ding:2019tqq,Shen:2022ffi,Beltran:2022ast,Zhou:2021ylt,Abada:2018sfh,Fernandez-Martinez:2022gsu,Abada:2022wvh,Arganda:2015ija,Bai:2022lbv,Das:2017nvm,Das:2015toa,Das:2016hof,Magill:2018jla,Izmaylov:2017lkv, Batell:2016zod, Bhattacherjee:2021rml,Accomando:2017qcs, Das:2019fee, Cheung:2021utb,Chiang:2019ajm, FileviezPerez:2020cgn,Das:2018tbd,Han:2021pun,Mason:2019okp, Accomando:2016rpc,Gao:2019tio,Gago:2015vma,Jones-Perez:2019plk}. 

The most considered production channel of the heavy neutrinos~($N$), are the charged-current Drell-Yan process~$pp \rightarrow W^{\pm (*)} \rightarrow l^{\pm} N$, with the subsequent decay $N \rightarrow l^{\mp} W^{\pm}$, and the $W$ dominantly decays into dijets. In the type-I seesaw, one can be benefited from the Majorana nature of the heavy neutrinos, leading to the "smoking-gun" signal of same-sign dilepton,~$l^{\pm} l^{\pm}$, which is considered to be almost SM background-free~\cite{Deppisch:2015qwa}. However, the heavy neutrinos and singlet $S$ are mass degenerate in the inverse seesaw, forming a pseuso-Dirac fermion pairs, while the Majorana component are highly suppressed. Therefore, we rely on the invariant masses of the lepton and dijets final states to search for the heavy neutrinos, which can be severely contaminated by the SM processes. 

This motivates the introduction of new techniques, such as 
the jet substructure techniques, which can be a powerful tool to help to resolve the problems. As considered in Ref.~\cite{Das:2017gke, Bhardwaj:2018lma,Chakraborty:2018khw}, the heavy neutrinos can lead to collimated topology of the dijets, forming a so-called "fat-jet" system. This can be of great help to improve significance and mitigating background. More recently, machine learning~(ML) techniques are also employed in Ref.~\cite{Feng:2021eke,Chakraborty:2018khw} to improve the sensitivity on the active-sterile mixing $V_{lN}$, but for trilepton plus missing energy final state. Comparing to the traditional cut-based analysis, ML might be more powerful to optimize the significance, especially if the system have large background. Since the hadronic final states have so, using the observables of the fat-jet system, we expect the ML methods can improve the sensitivity on the $V_{lN}$ via them as well.

Therefore, following the above literature, we employ the ML methods to search for the heavy Dirac heavy neutrinos in the inverse seesaw framework, at multiple options of colliders, including the High-Luminosity LHC~(HL-LHC) 13 TeV, High-Energy LHC~(HE-LHC) 27 TeV and future 100 TeV hadron colliders, e.g. Future Circular Collider~(FCC-hh)~\cite{FCC:2018vvp} or Super Proton-Proton Collider~(SPPC)~\cite{Tang:2015qga}. 
We focus on the hadronic final states~($l^+ l^- j j$) where the dijets forming a fat-jet signature, from the heavy neutrinos with masses, 100~GeV~$<m_N<$~1~TeV.
For the ML method, we use Gradient Boosted Decision Tree~(GBDT)~\cite{Roe:2004na} or Multi-Layer Perceptron~(MLP)~\cite{Fukushima:2013b} technique in the Scikit-learn framework~\cite{scikit-learn}.
The multi-variate are taken from the observables including the fat-jet system. Traditional cut-based analyses are also employed, and the resulting sensitivity for the two methods are compared. 

This paper is different to Ref.~\cite{Bhardwaj:2018lma}, as we introduce the ML methods, and Ref.~\cite{Feng:2021eke}, as we focused on the fat-jet final states instead of trilepton. Actually, since the branching ratio for the hadronic decays of $W$ is larger than the leptonically ones, we expect this channel might be better to search for the heavy neutrinos. Since it is harder to distinguish the background from the signal for the hadronic final states using cut-based analyses, we further expect the ML methods can make a greater difference.

This paper is organised as follows: We begin by briefly reviewing the inverse seesaw model in Sec.~\ref{sec:model}. The detailed analyses for the final states of the heavy neutrinos are discussed in Sec.~\ref{sec:ana}, including the setup of the event generation, the observables of the fat-jet system, and the cut-based as well as the ML analyses. Using this as a basis, the sensitivity on the $(m_N, V_{lN}^2)$ is derived and compared for the two methods in Sec.~\ref{sec:sen}, followed by the conclusion in Sec.~\ref{sec:con}.

\section{Inverse Seesaw Model}
\label{sec:model}
In addition to the particle content of the SM, the inverse seesaw incorporates two sets of fermions $N_{\alpha}$ and $S_{\beta}$, where $\alpha, \beta$ are the flavor indices, which are omitted for simplicity. They are singlets under the SM and carry same lepton numbers $L(N) = L(S)=1$. We consider three generations of $N$ and $S$, with the minimal Yukawa Lagrangian writes, 
\begin{eqnarray}
-\mathcal{L}_{Y} \ = \ 
Y \bar{\ell} \Phi N 
+ M_{N} \bar{S} N + \frac{1}{2}\mu_{S}\bar{S} S^C
+{\rm H.c.} \; ,
\label{eqn:Lagrangian}
\end{eqnarray}
where $\ell$ and $\Phi$ are the SM lepton and Higgs doublet, respectively. And $M_N$ and $\mu_S$ are the Dirac and Majorana mass term, with $\mu_S$ being the only one parameter violating the lepton number by two units.
$\bar{S} \equiv S^T C^{-1}$ is the charge conjugate of $S$. After the spontaneous breaking of the EW symmetry, we obtain the mass term,
\begin{eqnarray}
-\mathcal{L}_{mass} \ &=& \ 
M_D \bar{\nu_L} N + M_N \bar{S} N + \frac{1}{2} \mu_S \bar{S}S^C+{\rm H.c.} \\ &\equiv& \frac{1}{2}
\begin{pmatrix}
\bar{\nu_L} & N & \bar{S}
\end{pmatrix}
\begin{pmatrix}
{\mathbf 0} & M_D & {\mathbf 0} \\
M_D^{\sf T} & {\mathbf 0} & M_N^{\sf T} \\
{\mathbf 0} & M_N & \mu_S
\end{pmatrix}
\begin{pmatrix}
\nu_L^C \\
N \\
S^C \\
\end{pmatrix}.
\label{eqn:mxmass}
\end{eqnarray}
$\mu_S$ can be small naturally in the 't Hooft sense~\cite{tHooft:1979rat}, since the lepton number symmetry can be restored and light neutrino are massless when $\mu_S \rightarrow 0$.
Therefore $\mu_S \ll M_D \ll M_N$, leading to the light neutrino mass directly proportional to the $\mu_S$ as,
\begin{eqnarray}
M_\nu  \ = \ M_DM_N^{-1} \mu_S \: (M_N^{-1})^{\sf T} M_D^{\sf T} + \mathcal{O}  (\mu_S^3)\, ,
\label{eqn:mnu}
\end{eqnarray}
which provides a natural explanation for the origin of the light neutrino masses. The $N$ and $S$ form quasi-Dirac pairs $N_i$ and $\bar{N_i}$~($i=$1,2,3) with, 
\begin{eqnarray}
M_{N_i}, M_{\bar{N_i}} = m_N \mp \mu_S/2.
\end{eqnarray}
Here we use $m_N$ to represent the masses of the heavy neutrinos, whereas $M_N$ aforementioned is the mass matrix of them.

The Majorana nature of the heavy neutrinos as in the type-I seesaw, can be restored if a large Majorana mass term $\mu_R \bar{N} N^C$ is presented in Eq.~\ref{eqn:mxmass}. This can lead to the same-sign dilepton signatures. Nevertheless, we focus on the minimal inverse seesaw scenario, with only the lepton number conserving signals.

\begin{figure}[htbp]
\centering
\includegraphics[width=0.49\textwidth]{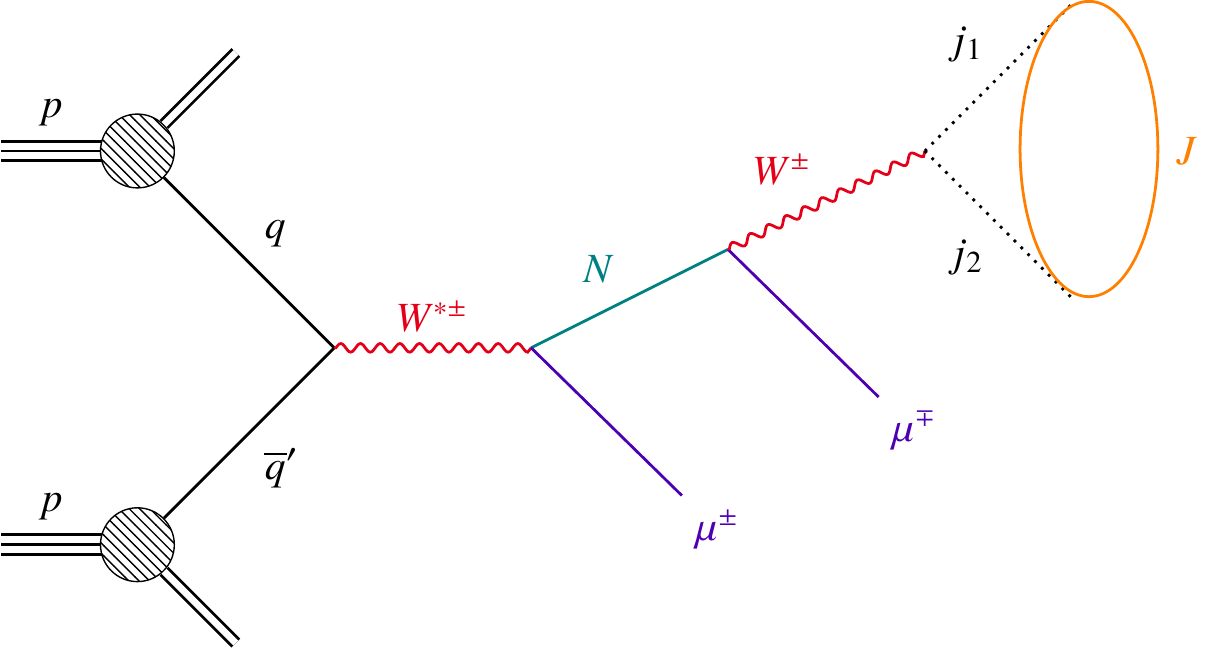}
\caption{The Feynman diagram of the process $pp \rightarrow W^{\pm *}\rightarrow \mu^{\pm} N \rightarrow \mu^{\pm} \mu^{\mp} W^{\pm} \rightarrow \mu^{\pm} \mu^{\mp} J$. The $W$ boson from the $N$ is boosted, so the hadronic final states of it become a fat-jet, $J$. }
\label{fig:feyn}
\end{figure}

For simplicity, we assume there is only one heavy neutrino relevant for the collider phenomenology while other heavy neutrino are too heavy thus not considered, and we omit $N_i$ as $N$. This heavy neutrino only interacts with the $\mu$ flavor, as it is detected with high efficiency and cleaner background.
Since we have sizeable active-sterile mixing, the primary production of the heavy neutrinos at the LHC are from the charged-current Drell-Yan process, and the heavy neutrinos dominantly decays into hadronic final states due to larger multiplicities, 
\begin{eqnarray}
pp \rightarrow W^{\pm *}\rightarrow \mu^{\pm} N \rightarrow \mu^{\pm} \mu^{\mp} W^{\pm} \rightarrow \mu^{\pm} \mu^{\mp} J,
\label{eq:process}
\end{eqnarray}
as their Feynman diagrams shown in Fig.~\ref{fig:feyn}.
This processes lead to two charged lepton plus two jets~(dijets) final states. We focus on the heavy neutrinos with masses 100~GeV~$< m_N <~$1~TeV, so the $N$ decays into $W$ on-shell, and $N$ is not so heavy that gluon fusion and vector-boson fusion becomes more important~\cite{Feng:2021eke}. Since the $N$ is heavier than the $Z,W$ and $H$ bosons in most cases, we simply have $Br(N\rightarrow l^{\pm} W^{\mp})$:$Br(N\rightarrow \nu Z)$:$Br(N\rightarrow \nu H)$=2:1:1, i.e. the Goldstone boson equivalence theorem. Therefore, $Br(N\rightarrow \mu^{\pm} W^{\mp}) \approx 50\%$.
Once $N$ is sufficiently large, the daughter $W$ is boosted, and the dijets in the final states becomes collimated, leading to a fat-jet signature, represented by the $J$ in Eq.~\ref{eq:process}. 

The signals contain two opposite sign di-leptons and a fat-jet. The former can also come from the gauge boson as well as $t \bar{t}$ decays from the SM processes. The fat-jet can be faked by additional QCD jets. Following Ref.~\cite{Bhardwaj:2018lma}, the leading background processes at hadron colliders are,
\begin{eqnarray}
pp &\rightarrow& t \bar{t} + j\rightarrow \mu^{\pm} \mu^{\mp} + MET + j+b+\bar{b}, \\
 &\rightarrow& W^{\pm} W^{\mp} + j\rightarrow \mu^{\pm} \mu^{\mp} + MET + j, 
\label{eq:processb}
\end{eqnarray}
where $MET$ represents the missing transverse energy, which is produced via the missing neutrinos, as well as charged leptons. 
The $Z$ boson mediated background, $pp \rightarrow Z + j \rightarrow \mu^{+} \mu^{-}+ j$ and $\rightarrow Z W^{\pm} + j \rightarrow \mu^{+} \mu^{-}+  jjj$ is cut out by a pre-cut $M_{\mu^+ \mu^-} >$ 200 GeV.

Based on that, in order to separate the signal from the background, we are going to describe the detailed analyses in the following section. Towards the different objects in the final states, cuts are implemented for cut-based analyses, or the corresponding observables are fed to the ML analyses. 

\section{Analysis}
\label{sec:ana}
To obtain the information of the final states at colliders for later analyses, we simulate the signal and background events in the following steps.
We use the Universal FeynRules Output (UFO)~\cite{Degrande:2011ua} of the heavy neutrino model provided in Ref.~\cite{Degrande:2016aje}, which is fed to the Monte Carlo event generator {\tt MadGraph5aMC@NLO}~\cite{Alwall:2014hca} to generate parton level events, with {\tt NN23LO1}~\cite{NNPDF:2014otw} parton distribution function. The initial and final state parton shower, hadronization, heavy hadron decays, etc are performed by {\tt PYTHIA 8}~\cite{Sjostrand:2014zea}. The shower jets and the matrix element jets are matched by MLM matching~\cite{Mangano:2006rw,Hoeche:2005vzu}. Matched background is generated using the default kt-MLM algorithm with {\tt Xqcut} = 30 GeV and the jet matching parameter~(QCUT) 1.5 times larger than the {\tt Xqcut}. The clustering of events is performed by {\tt FastJet}~\cite{Cacciari:2011ma}, and the detector level simulation is implemented by {\tt Delphes}~\cite{deFavereau:2013fsa}.

\begin{figure}[htbp]
\centering
\includegraphics[width=0.8\textwidth]{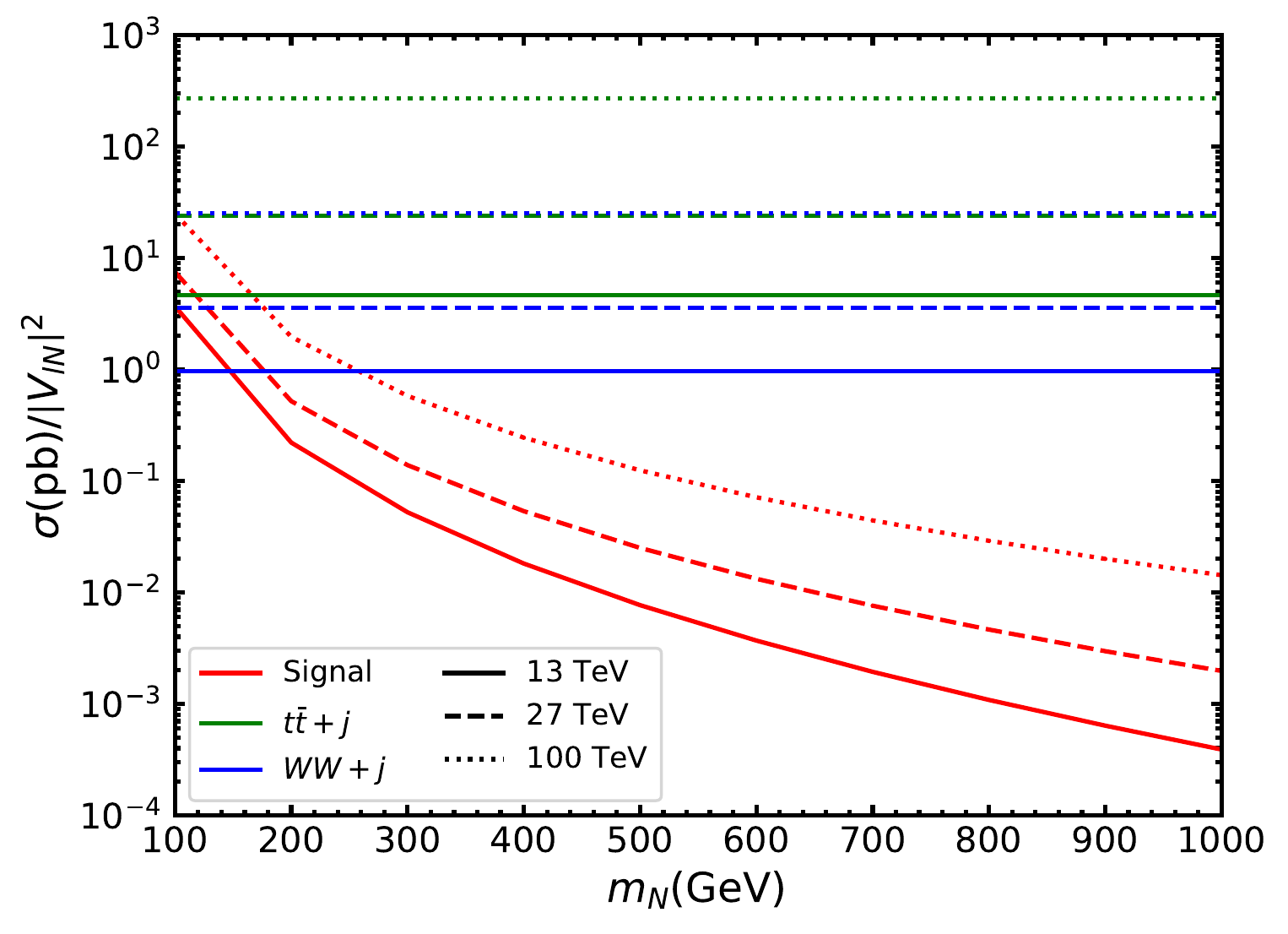}
\caption{The cross section of the signal process, $pp \rightarrow W^{\pm *}\rightarrow l^{\pm} N \rightarrow l^{\pm} l^{\mp} W^{\pm} \rightarrow l^{\pm} l^{\mp} J$ and the two leading background processes, $pp \rightarrow t \bar{t} j\rightarrow l^{\pm} l^{\mp} + MET + j+b+\bar{b}$ and $pp \rightarrow W^{\pm} W^{\mp} j\rightarrow l^{\pm} l^{\mp} + MET + j$ at the 13, 27, 100 TeV hadron colliders.}
\label{fig:pro}
\end{figure}

The resulting production cross section of the signal and leading background processes at 13, 27, 100 TeV hadron colliders is shown in Fig.~\ref{fig:pro}. NNLO effects are considered, as we apply a k factor of two for the signal~\cite{Cottin:2022nwp}, and the background cross section is normalized according to Ref.~\cite{Bhardwaj:2018lma}.
The cross section of the signal process can reach $\mathcal{O}$($1$)~pb when $V_{lN}^2=$1, while the background processes approach similar magnitude at 13 TeV. Increasing the collision energy from 13 to 100 TeV, enlarges the cross section of the signal and background for roughly one magnitude.
Since at the EWPD limits $V_{lN}^2<10^{-3}$~\cite{delAguila:2008pw,Akhmedov:2013hec,Antusch:2014woa,Blennow:2016jkn}, there is a big gap between the number of events from the signal and background. Thus, optimized cuts are required to separate them. 

\subsection{Cut Based Analysis}
The final states of the signal and background processes contain muons, b-jets, missing energy and the fat-jet. Thus, the observables of our primary interests includes the $p_T$, $\eta$ of the muons and fat-jet, the b-tag of the jets and the MET. To remove the background from boson decays, i.e. $Z+j$ and $Z+W+j$ processes, the invariant masses of the dimuon system $M(\mu^+ \mu^-)$ should also be useful. We can extract more information from the jet substitute, like the invariant masses of the fat-jet, and the N-subjettiness ratio $\tau_{21} = \tau_2/\tau_1$ for the two component of the fat-jet, as defined as~\cite{Thaler:2010tr,Thaler:2011gf},
\begin{eqnarray}
 \tau_N^{(\beta)} = \frac{1}{\mathcal{N}_0} \sum\limits_i p_{i,T} \min \left\lbrace \Delta R _{i1}^\beta, \Delta R _{i2}^\beta, \cdots, \Delta R _{iN}^\beta \right\rbrace,
 \label{eq:nsub_N}
\end{eqnarray}
where $\mathcal{N}_0=\sum\limits_i p_{i,T} R_0$ with jet radius $R_0$, the radius between two object is $ \Delta R_{i\alpha} = \sqrt{(\Delta \eta)^2_{i\alpha}+(\Delta \phi)^2_{i\alpha}}$, and the thrust measure $\beta = 2$ as we choose. The $t \bar{t}$ background can be reduced if the transverse mass variable $M_{T2}$~\cite{Cao:2017ffm,Cheng:2008hk} of the $(\mu^+ \mu^- J)$ system is considered. 
Since the heavy neutrinos decay into $\mu^\pm  J$, the invariant masses of the two particles, $M(\mu^\pm J)$ should be able to reconstruct the $N$ masses, which is also helpful to identify the signal.

\begin{table}
	\centering
	\begin{tabular}{|l|c|c|c|c|c|c|c|}
		\hline
		Final states & $p_T(\mu_{1,2})$ & $|\eta(\mu^\pm)|$ & $M(\mu^+ \mu^-)$ & b-tag & MET & $p_T(J)$ \ \\
		\hline
		$\mu^+ \mu^- J$ & >100,60~GeV &  <2.4 & <200~GeV  & 0 & <60~GeV & >150~GeV\\
		\hline
    $|\eta(J)|$ &$M^{J}$ & $\tau_{21}(J)$ & $M_{T2}^{\mu^+ \mu^- J}$ & $M(\mu^\pm J)$ & $M_{T}^{\mu^\pm J}$ & \\  \hline
      <2.4 & >50~GeV & <0.4 & > 250~GeV & > 500~GeV & < 60 GeV & \\ \hline
	\end{tabular}
	\caption{The cut-based analysis, following Ref~\cite{Bhardwaj:2018lma}, with an additional cut on $M_{T}^{\mu^\pm J}$.}
	\label{tab:cut}
\end{table}
\begin{figure}[htbp]
\centering
\includegraphics[width=0.8\textwidth]{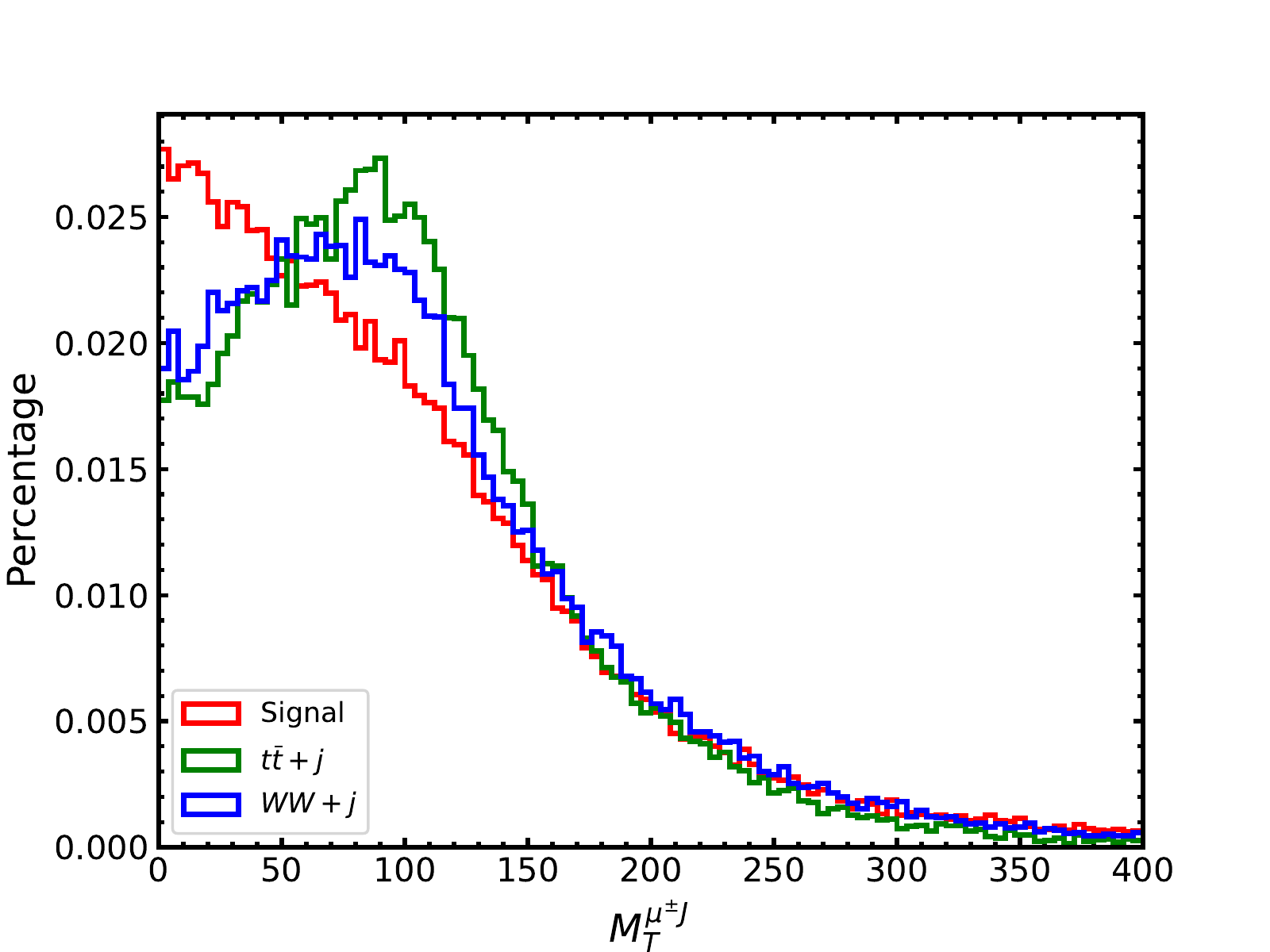}

\caption{$M_{T}^{\mu^\pm J}$ distributions of the signal and background processes, with $m_N$ = 400 GeV at 13 TeV LHC.}
\label{fig:MTmj}
\end{figure}

Following Ref.~\cite{Bhardwaj:2018lma}, a cut-based analysis~(CBA) can be implemented as shown in Table.~\ref{tab:cut}, with an additional cut on $M_{T}^{\mu^\pm J}$. We require $M_{T}^{\mu^\pm J} <$ 60 GeV to separate the background from signal, as suggested in Fig.~\ref{fig:MTmj}.
Here the final states requirements include base-line cuts, as we ask for exactly two opposite muons with $p_T(\mu)>$ 20 GeV and $|\eta(\mu)|<$ 2.4, and at least one fat-jet with radius parameter $R=$ 0.8, $|\eta(J)|<$ 2.4, and the leading fat-jet has $p_T(J)>$ 100 GeV. The cuts on the fat-jet in Table~\ref{tab:cut} is always put on the leading fat-jet.

\subsection{Machine Learning}
Machine learning is widely used to recognize events generated by different processes. For different kinds of data objects, different representations can be used to pass them to the neural network for inference learning, for example image representation is intended to represent the signal received by the detector from a particle as a picture of multiple channels, with information related to the particle such as transverse momentum, PDG ID, charge, etc., that can be stored in different channels of information for the same pixel. The graph is a more natural way of representing each event abstractly as a particle-to-particle association, and usually such an association can be represented by $\triangle R$. This way it is convenient to add not only a description of the particle but also a description of the behavior of the whole, e.g. N-subjettiness. In this paper, well-defined observables are used to describe events, compared with the use of cut-based method, gradient boosted decision tree (GBDT), and multilayer perceptron (MLP).

We use GBDT provided by Sklearn\cite{scikit-learn} package under all the default parameters. The MLP model is built by Keras\cite{chollet2015keras}, a high level interface of the machine learning framework Tensorflow\cite{tensorflow2015-whitepaper}. To capture all potential hidden information underneath observables, we build a MLP with ten layers. Its structure is FC16-FC32-FC64-FC128-FC64-FC32-DP-FC16-DP-FC2, where FC is for a fully connected layer (Dense layer in Keras), DP is for dropout\cite{JMLR:v15:srivastava14a} layer with drop rate 0.5 to prevent the model from overfitting. We choose Adam\cite{kingma2017adam} as the optimizer and sparse cross-entropy as the loss function. We simulate 150,000 valid colliding events after pre-cuts and split them into train/test dataset with the ration 7:3.

Using these two ML methods, we expect the new analyses can help to separate the background from signal. To employ the ML techniques, we need to input the signal and background, and their observables. These observables in principle, should already be able to identify the signal and background. From Table~\ref{tab:cut}, we have 11 observables in total for the CBA analyses. The distribution of these observables for signal and background are already shown in Ref.~\cite{Bhardwaj:2018lma}, here we only show the distribution for the additional one, $M_{T}^{\mu^\pm J}$ in Fig.~\ref{fig:MTmj}, with $m_N=$ 200~(left), 800~(right) GeV. From these distribution, all observables except the $\eta$ of jets and muons, have shown significant difference between the signal and background, hence can be inputs to the ML analyses.

Therefore, we classify the 11 observables left, into three categories mainly based on their particles, 
\begin{itemize}
    \item 3 lepton observables, $p_T(\mu_{1,2})$,  \text{MET}
    \item 3 jet observables, $p_T(J)$, \text{b-tag}, $\tau_{21}(J)$
    \item 5 mass observables, $M(\mu^+ \mu^-)$, $M^{J}$, $M_{T_2}^{\mu^+ \mu^- J}$, $M(\mu^\pm J)$, $M_{T}^{\mu^\pm J}$.
    \end{itemize}

\begin{figure}[htbp]
\centering
\includegraphics[width=0.495\textwidth]{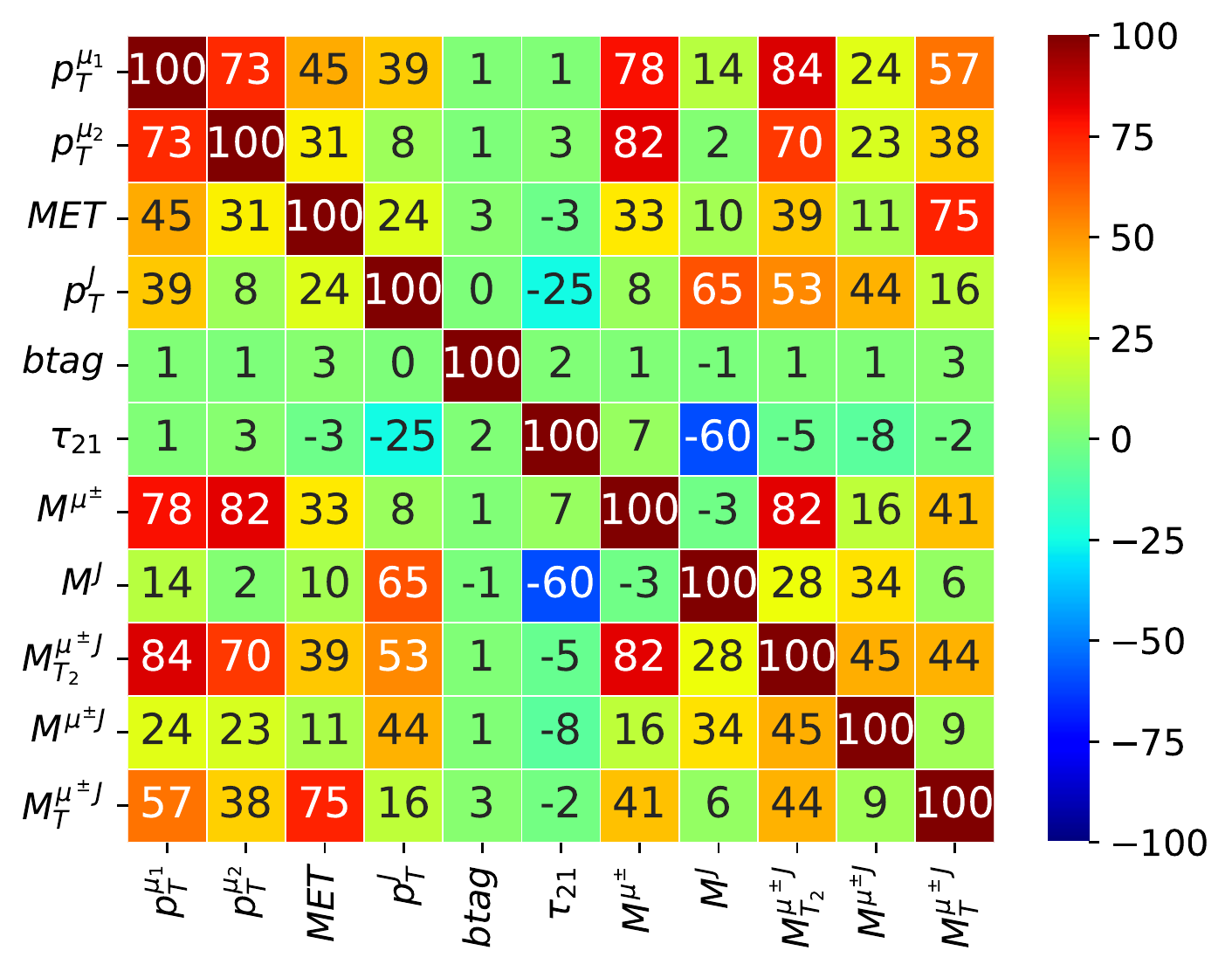}
\includegraphics[width=0.495\textwidth]{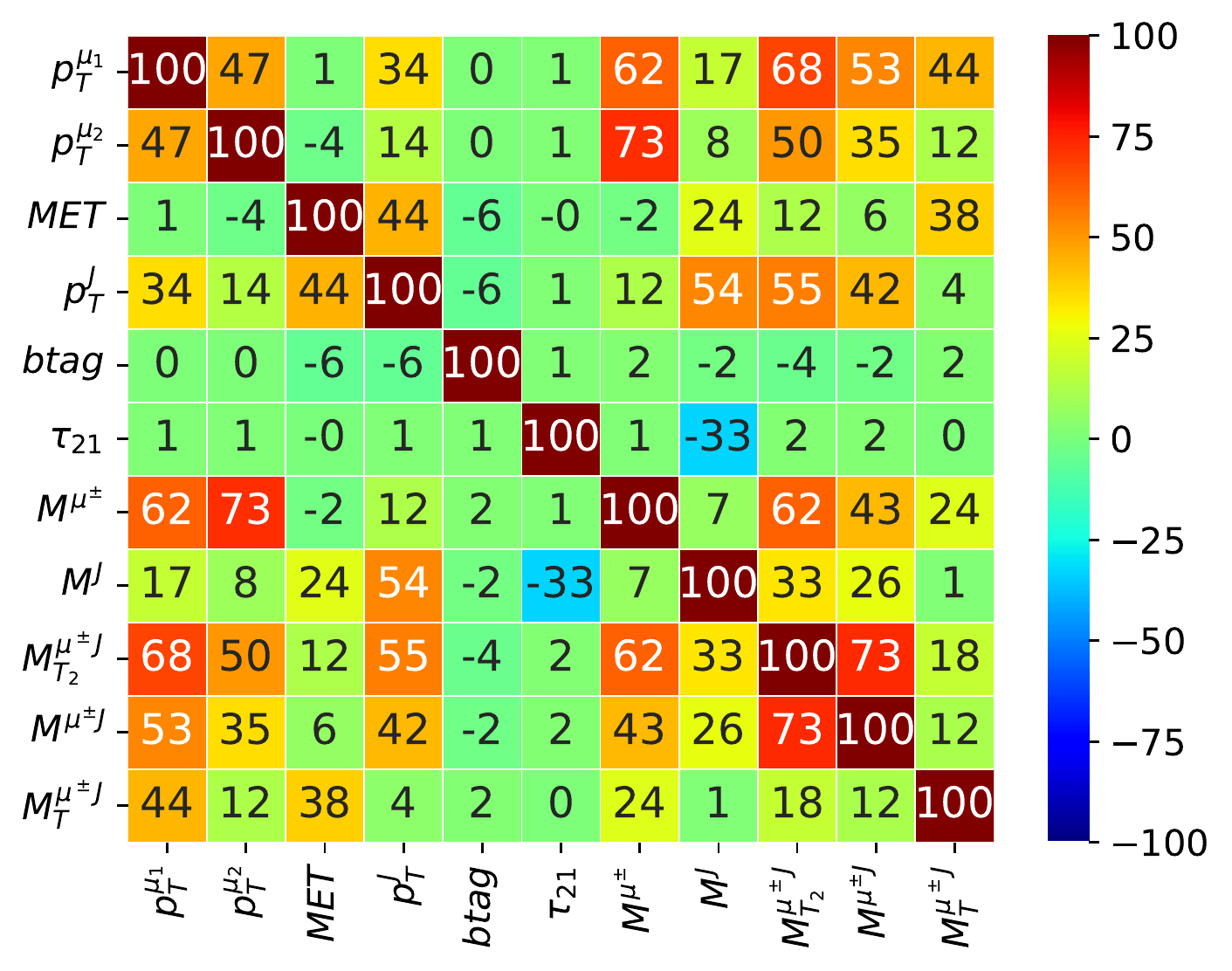}
\caption{Correlation matrix for signal~(left) and background~(right) at 13 TeV, when $m_N$ = 400 GeV.}
\label{fig:co_sg}
\end{figure}

These observables are not independent, and we show their linear correlation matrix for signal~(top) and background~(bottom) at 13 TeV, when $m_N$ = 400 GeV in Fig.~\ref{fig:co_sg}. In general, the observables for the background are more independent from others.
Among the obervables, the three lepton ones are correlated in a relatively large degree. The mass observables are highly dependent on the transverse momentum of their components. The transverse momentum of the fat-jet is correlated to the ones from the lepton and missing energy.
Anyhow, the b-tag and N-subjettiness ratio $\tau_{21}$ seems to be independent on any other observables, severed as the most unique ones. Afterwards, this correlation matrix is used by the training processes of the ML methods.

\begin{figure}[htbp!]
\centering
\includegraphics[width=0.49\textwidth]{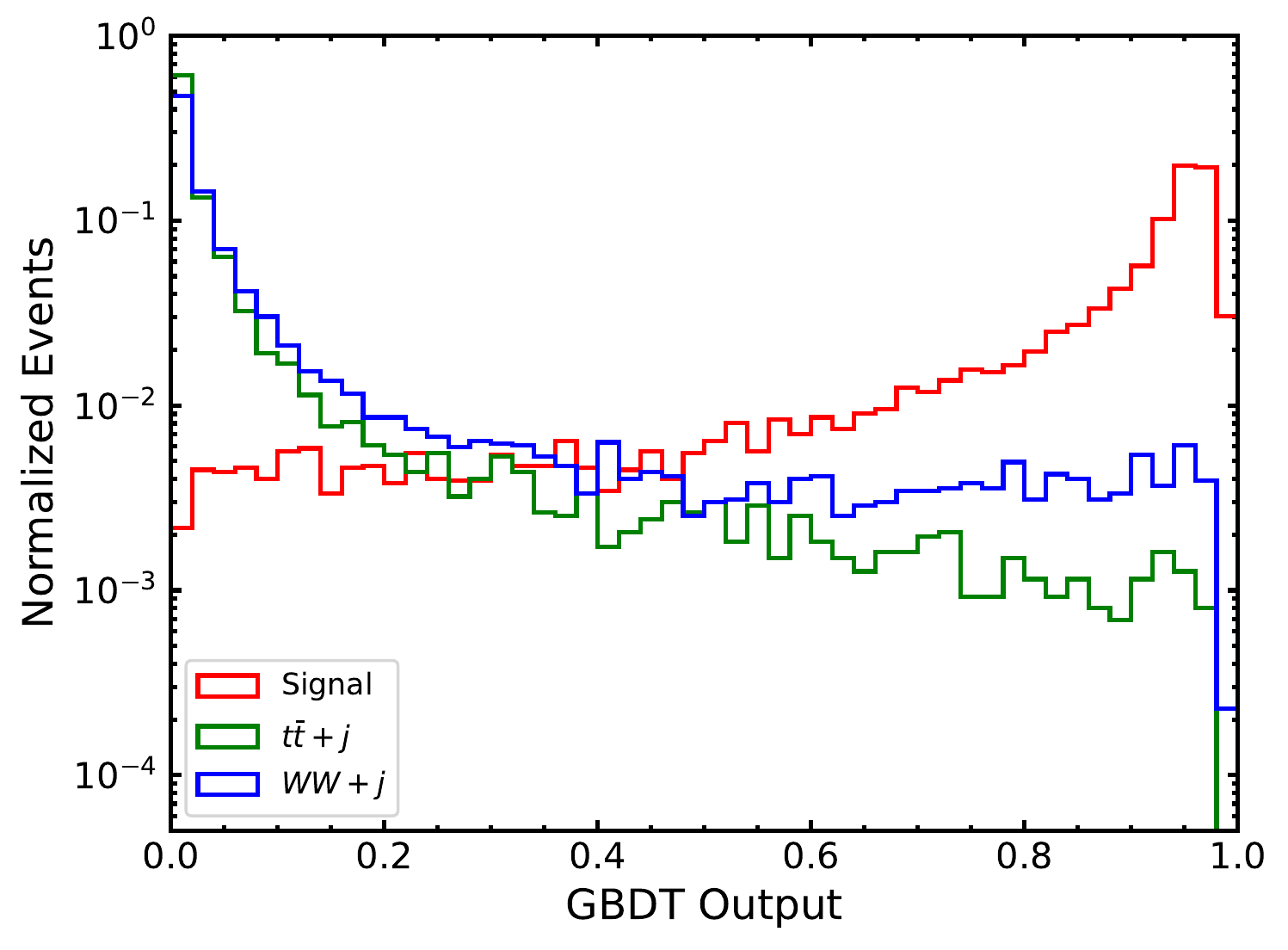}
\includegraphics[width=0.49\textwidth]{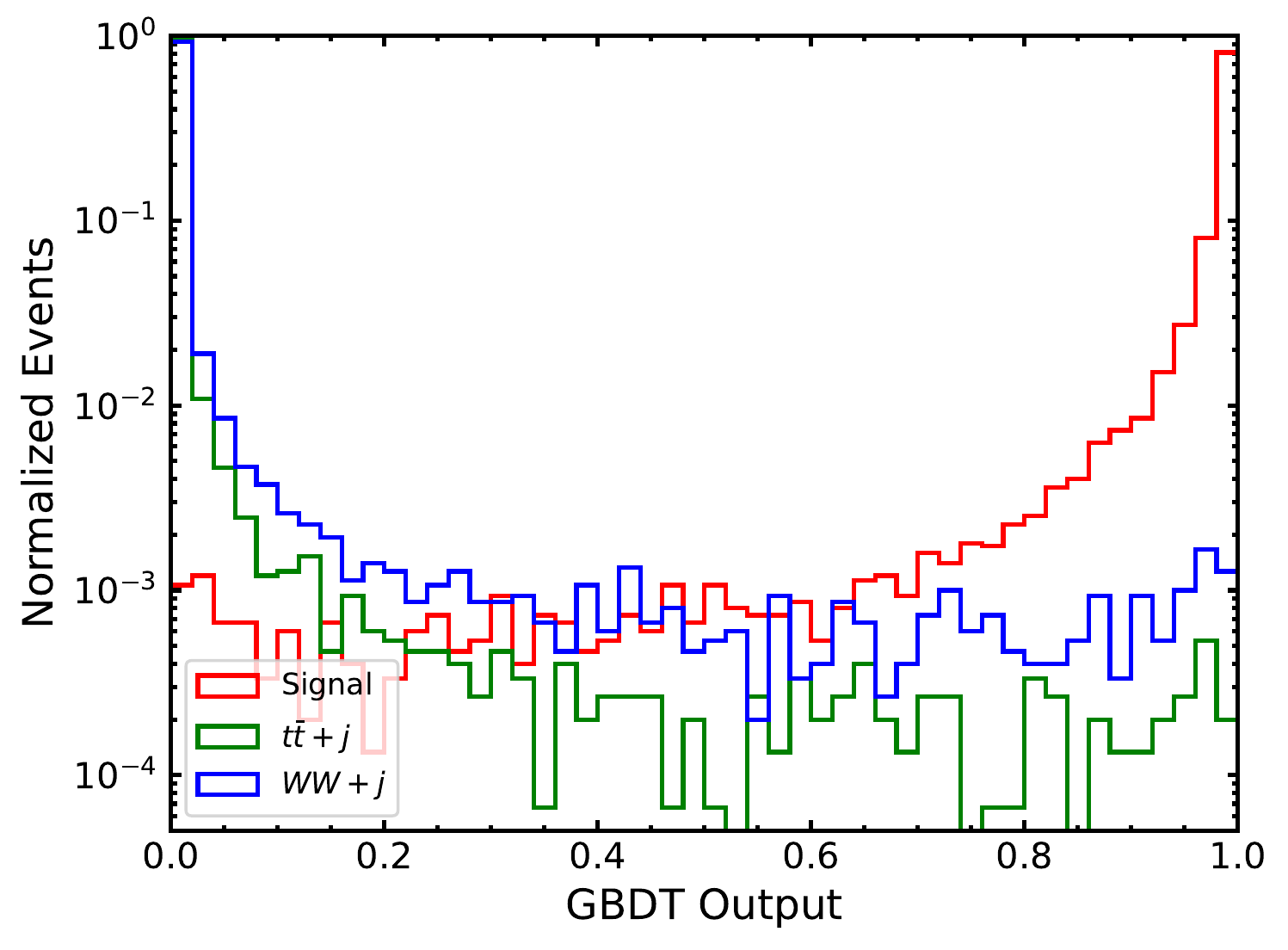}
\includegraphics[width=0.49\textwidth]{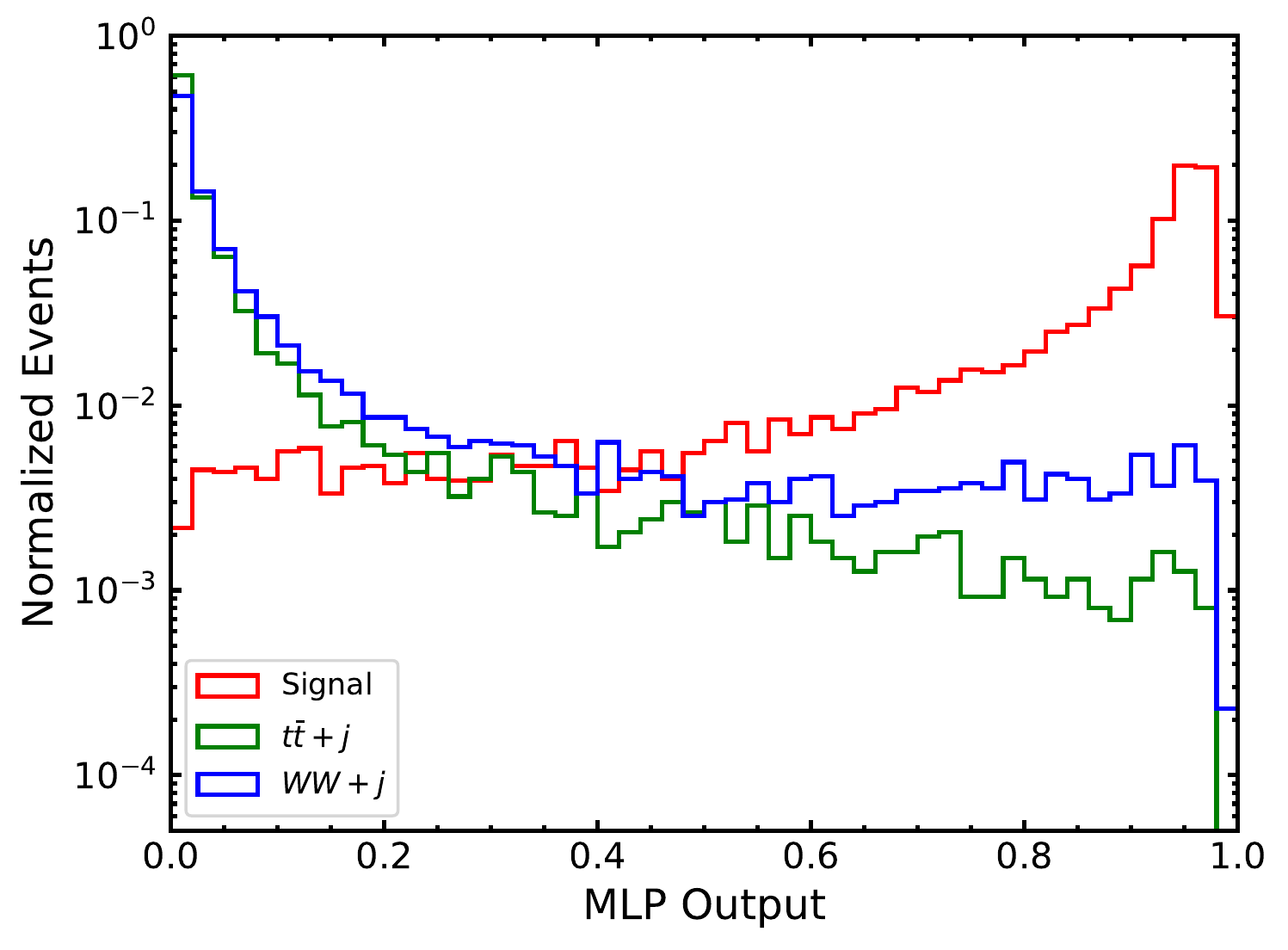}
\includegraphics[width=0.49\textwidth]{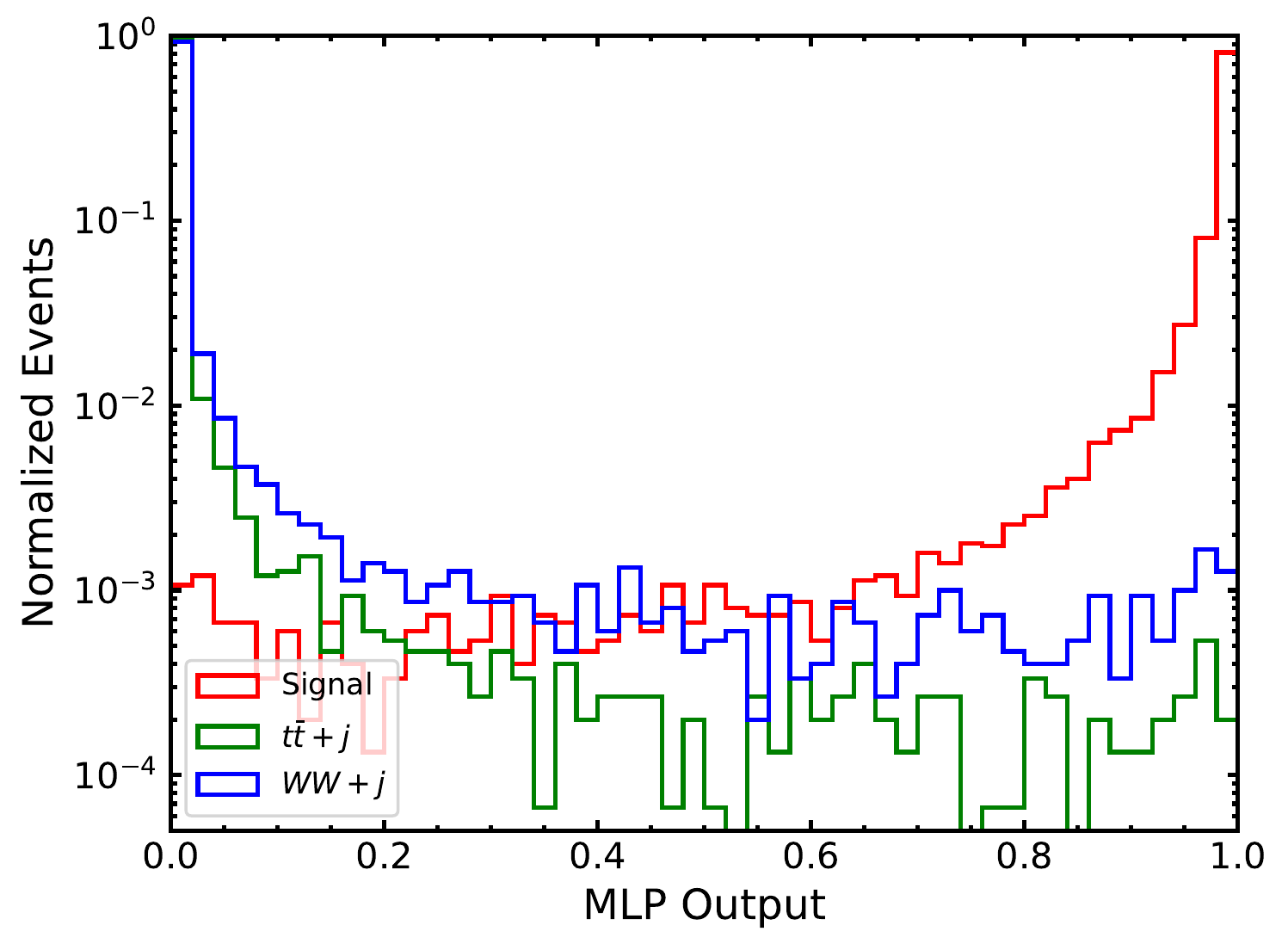}

\caption{ML estimator distributions of the signal and background processes, with $m_N$ = 200~(left) and 800~(right) GeV at 13 TeV LHC.}
\label{fig:ML_est}
\end{figure}
After inputting the observables of the background and signal, by performing the two ML methods, we obtain a single likelihood variable, i.e. the ML estimator, as shown in Fig.~\ref{fig:ML_est}, for GBDT~(top) and MLP~(bottom) respectively, when $m_N=$ 200~(left), 800~(right) GeV. In the figures, when the $x$ axis moves from 0 to 1, the likelihood of the event which is identified as a signal increases. Overall, both two methods are powerful to distinguish signal from background, as the background events which are misidentified as signal are at least $\mathcal{O}(10^2)$ smaller than the right ones. The two methods have shown very similar distribution, thus ability. Comparing the two background processes, $t \bar{t} + j$ and $WW + j$, the former is better identified, but only in a small degree. When the masses of the heavy neutrinos increase from 200 to 800 GeV, it becomes easier to distinguish the background from signal. This can be understood, as the signal events now contain a relatively heavy fat-jet, making the jet and mass observables significantly different between signal and background.

\begin{figure}[htbp!]
\centering
\includegraphics[width=0.49\textwidth]{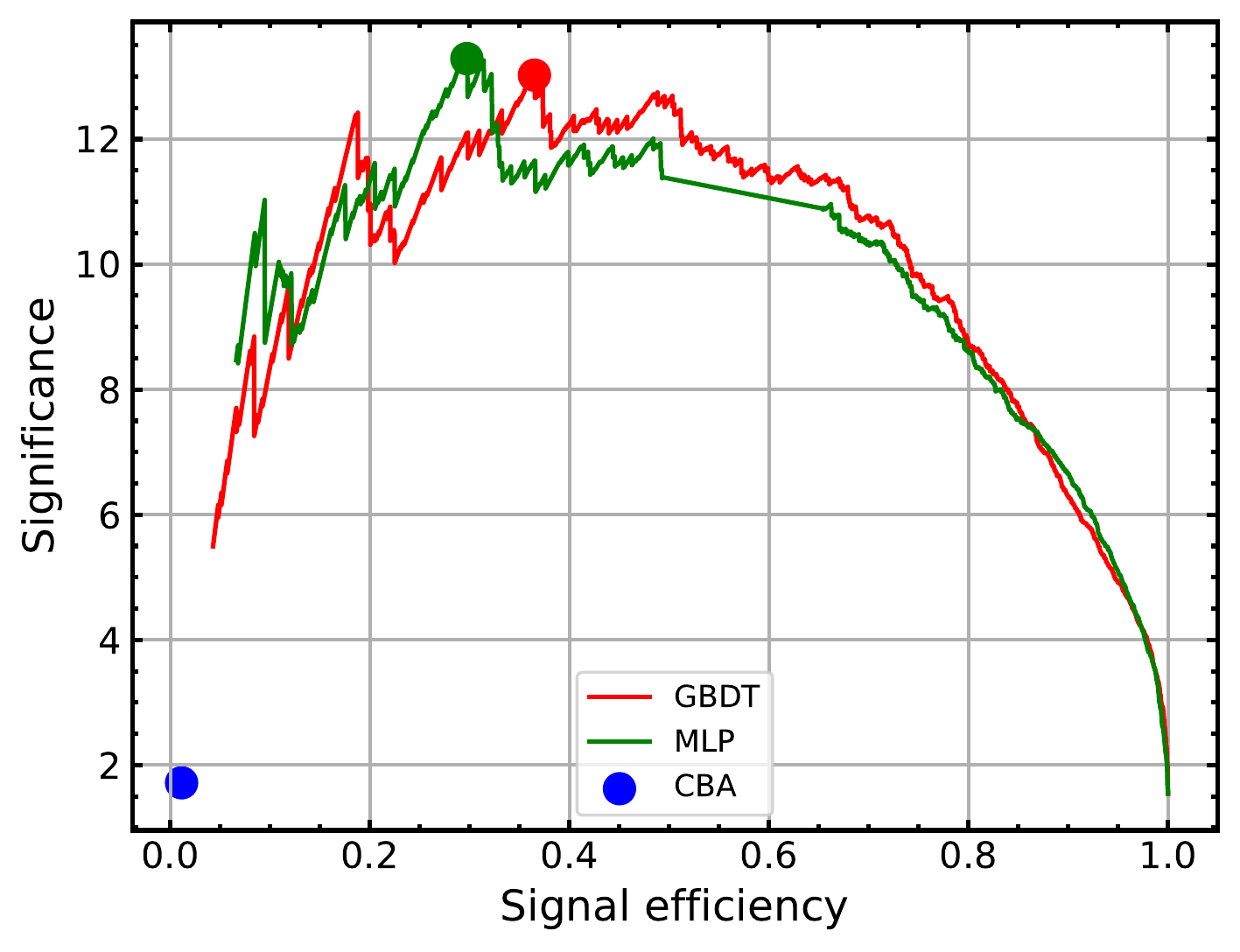}
\includegraphics[width=0.49\textwidth]{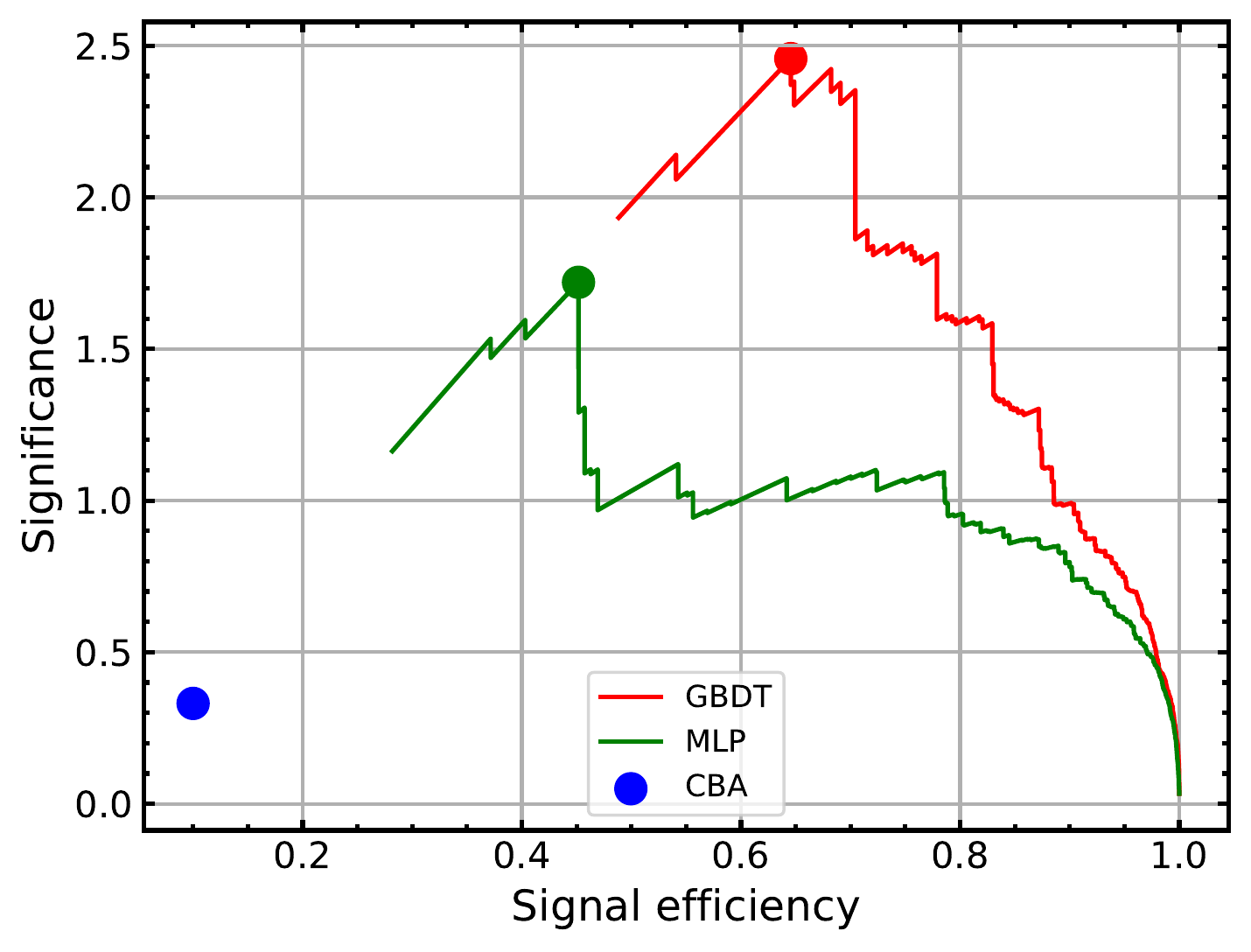}
\includegraphics[width=0.49\textwidth]{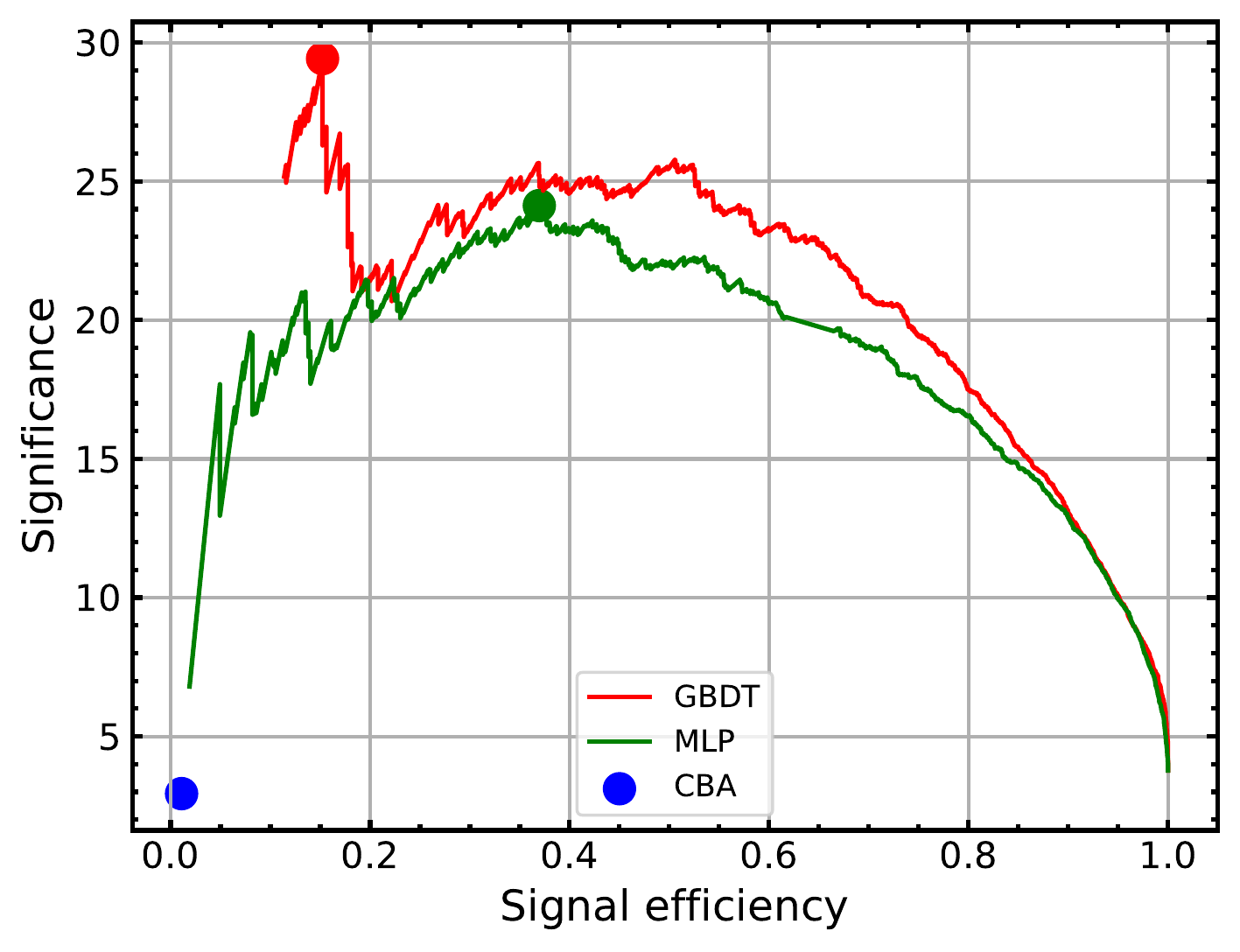}
\includegraphics[width=0.49\textwidth]{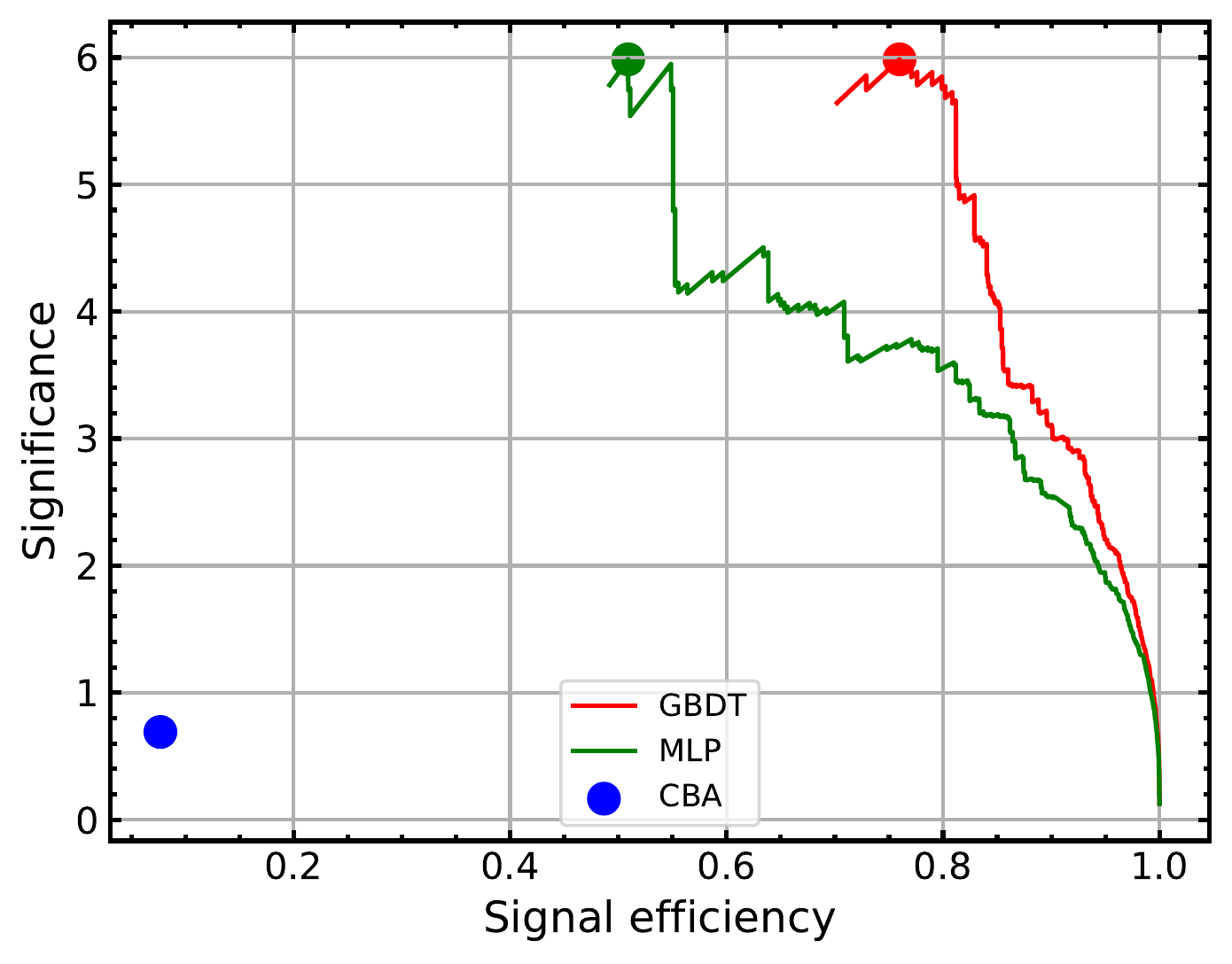}
\includegraphics[width=0.49\textwidth]{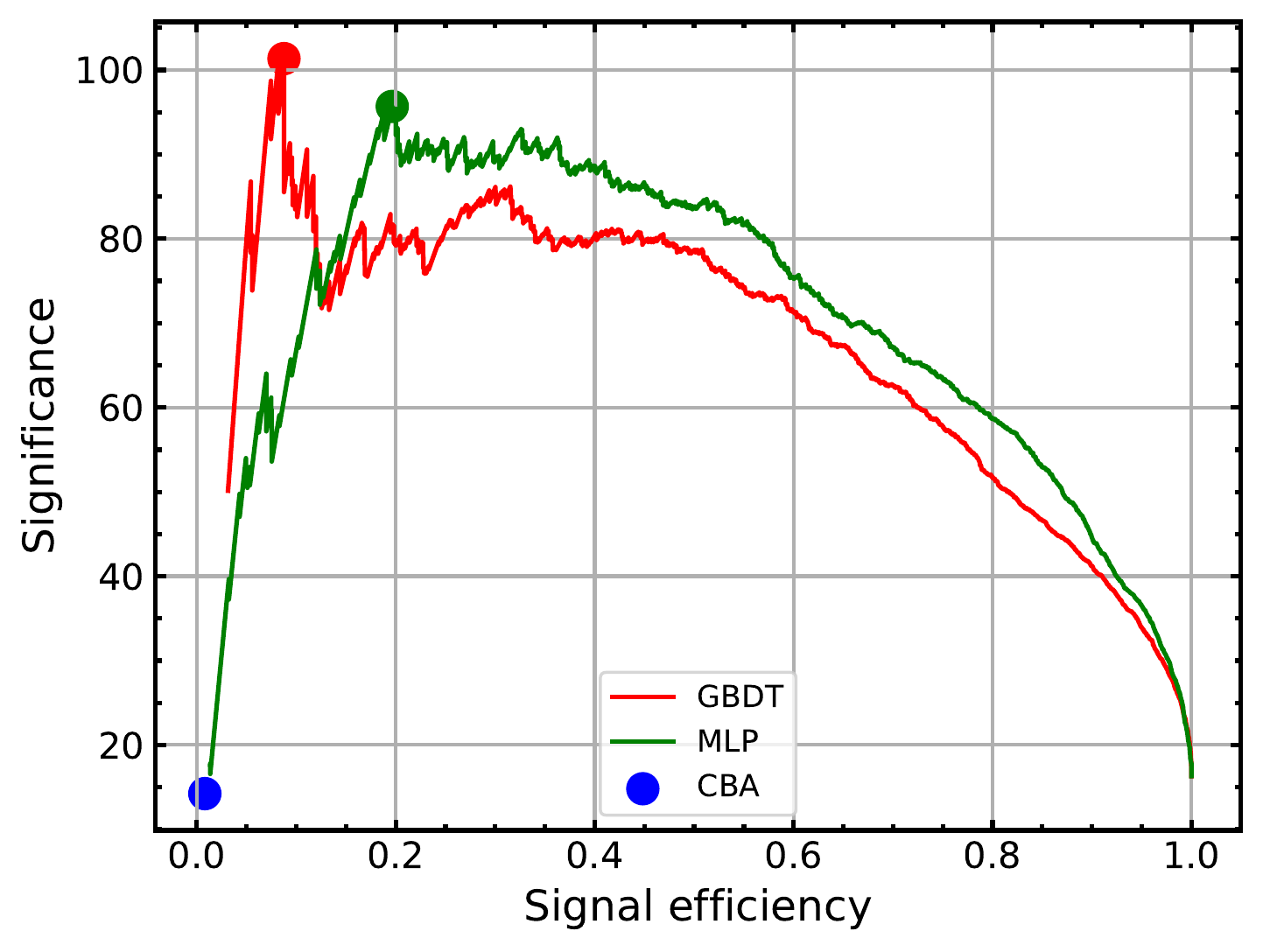}
\includegraphics[width=0.49\textwidth]{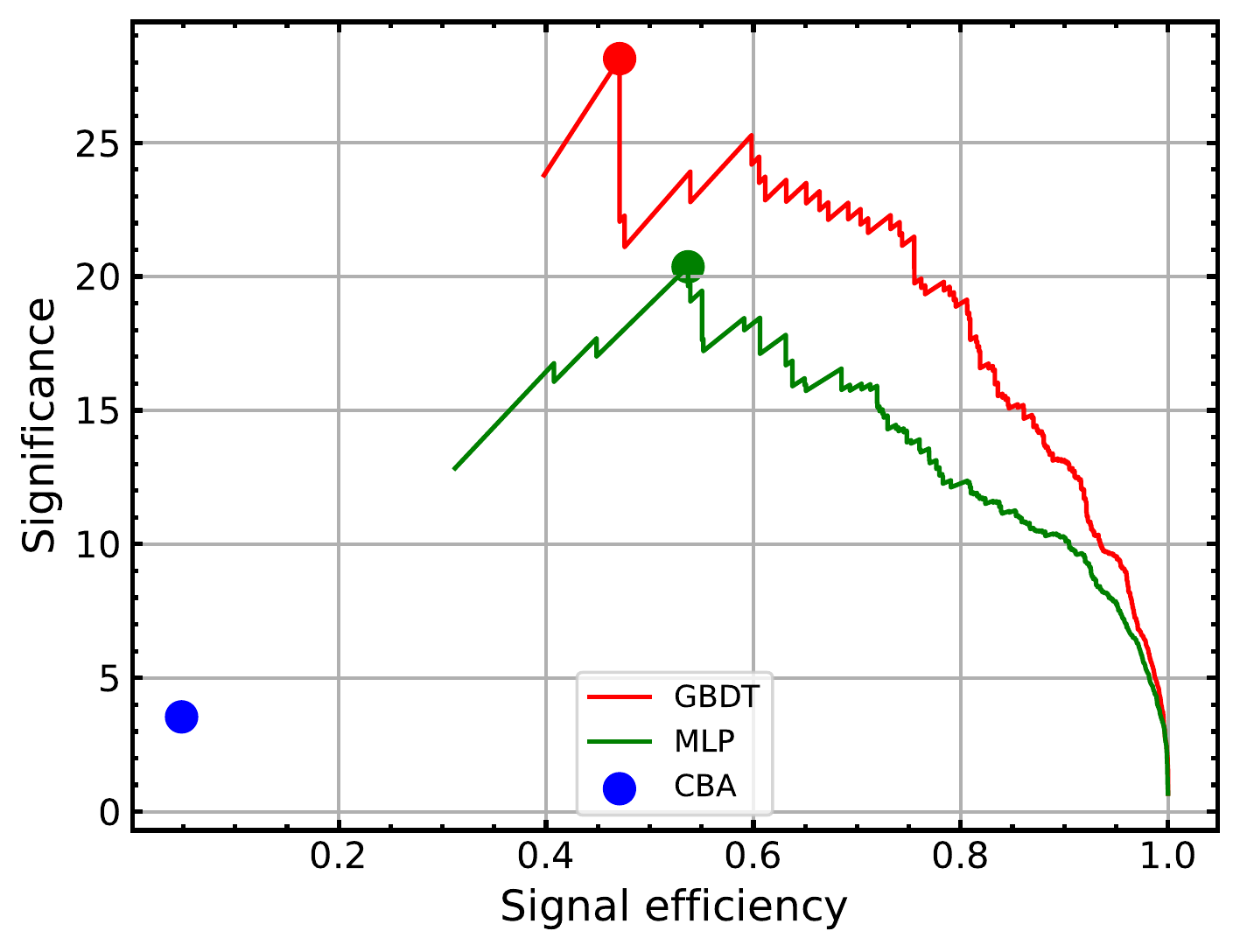}
\caption{Significance versus signal efficiency at the 13~(up), 27~(middle), 100~(bottom) TeV hadron colliders, for $m_N$ = 200~(left), 800~(right) GeV. We fix $V_{\mu N}^2$ = 1.}
\label{fig:ML_bs}
\end{figure}

With the ML estimator in hand, we can now derive the background rejection and signal efficiency, and the resulting significance, $S/\sqrt{B}$. Depending on the ML estimator value we choose, the signal efficiency and the corresponding background rejection are different. In general, if we want to achieve the best background rejection, we usually left with small signal efficiency, and vice versa. Combing the two effects, so the largest significance is obtained when the signal efficiency is in the middle.
The distribution of significance versus the signal efficiency is shown in Fig.~\ref{fig:ML_bs}, at the 13~(up), 27~(middle), 100~(bottom) TeV hadron colliders, for $m_N$ = 200~(left), 800~(right) GeV, and we fix $V_{\mu N}^2$ = 1. The significance for other values of $V_{\mu N}^2$ can be obtained by a simple re-scaling.
From the distribution, we can see that GBDT has better potential to distinguish the background from signal in general. To achieve the same signal efficiency, the background rejection for GBDT is larger than the MLP. From the distribution, we can optimize the ML estimator value, to yield the best significance. The points where we can get best significance is illustrated in Fig.~\ref{fig:ML_bs}, and we will use these points to get the sensitivity for the $V_{\mu N}^2$ as a function of $m_N$ in the following section. The points of CBA analyses are also put for comparison. They can lead to sufficiently large background rejection. However, the signal efficiency is significantly smaller than the ML methods, giving rise to much smaller significance.

\begin{figure}[htbp!]
\centering
\includegraphics[width=0.49\textwidth]{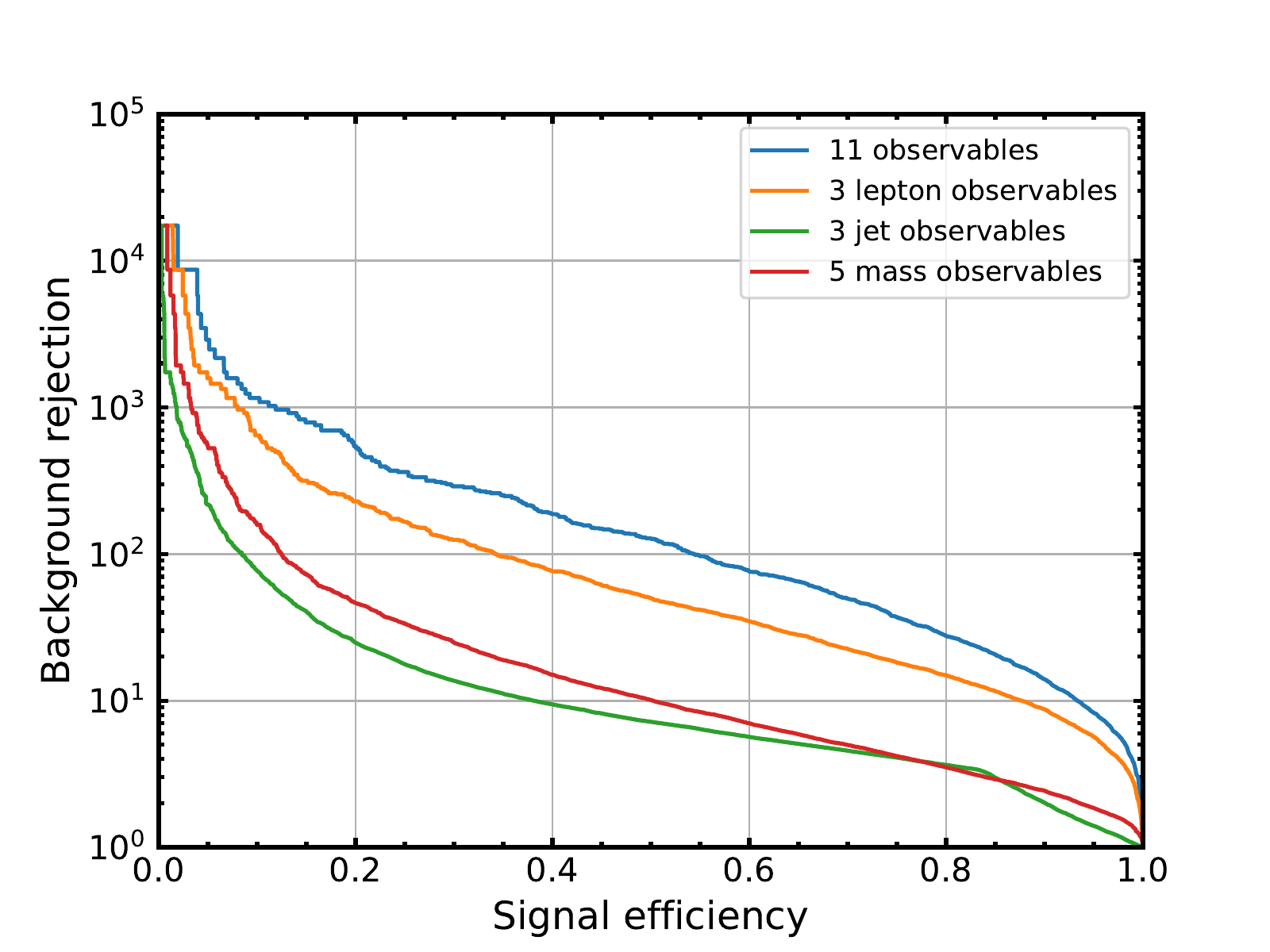}
\includegraphics[width=0.49\textwidth]{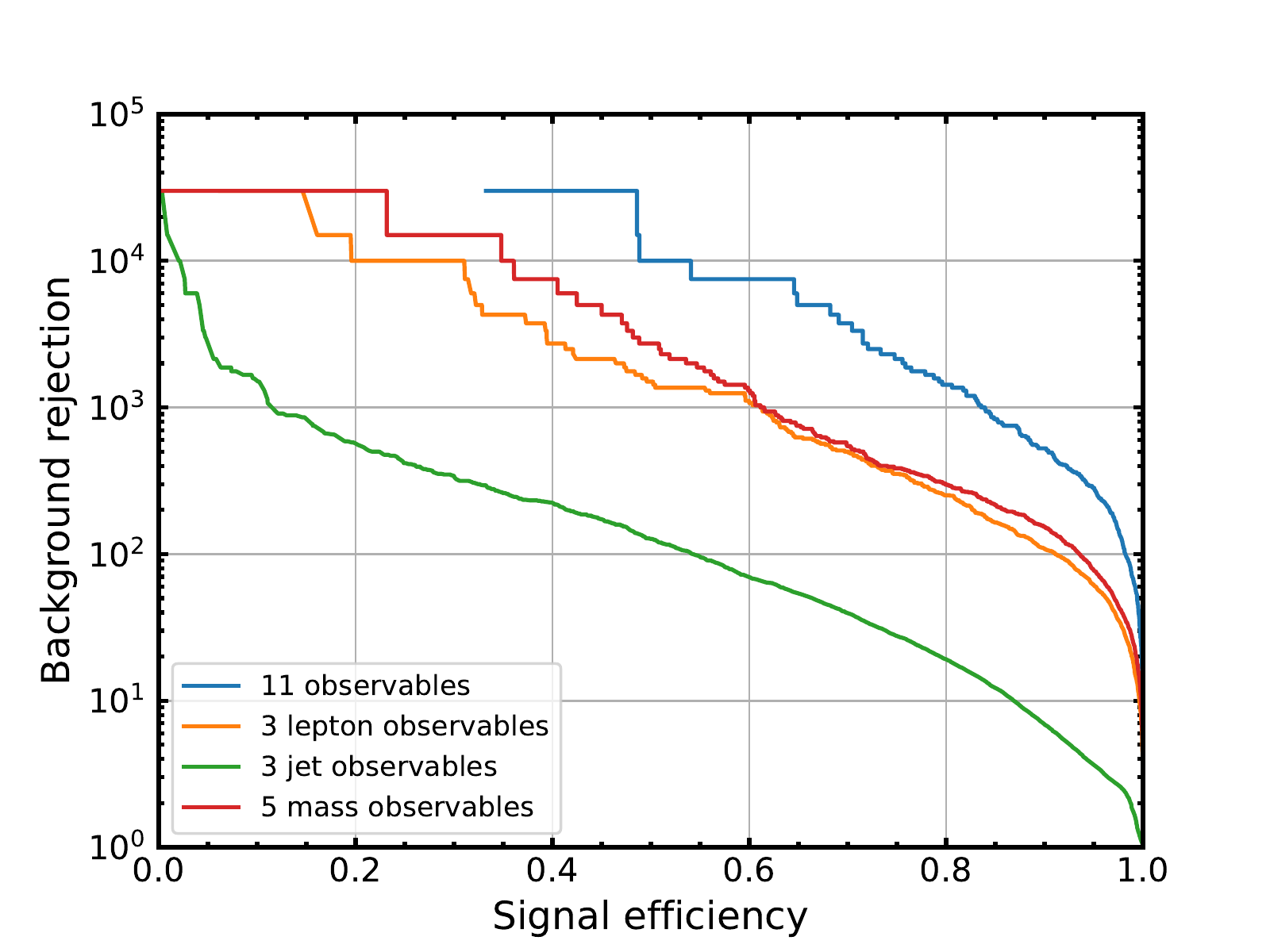}
\includegraphics[width=0.49\textwidth]{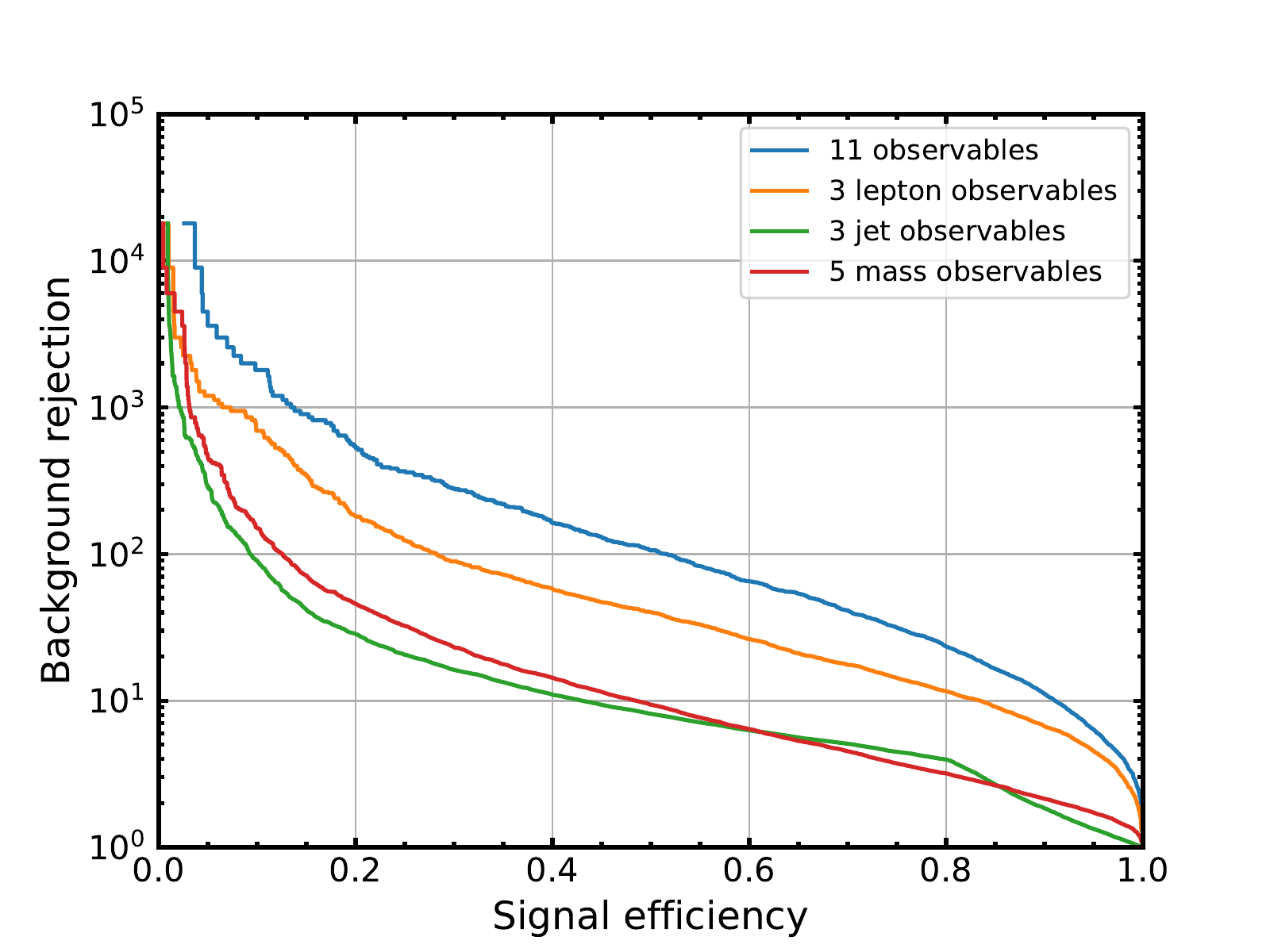}
\includegraphics[width=0.49\textwidth]{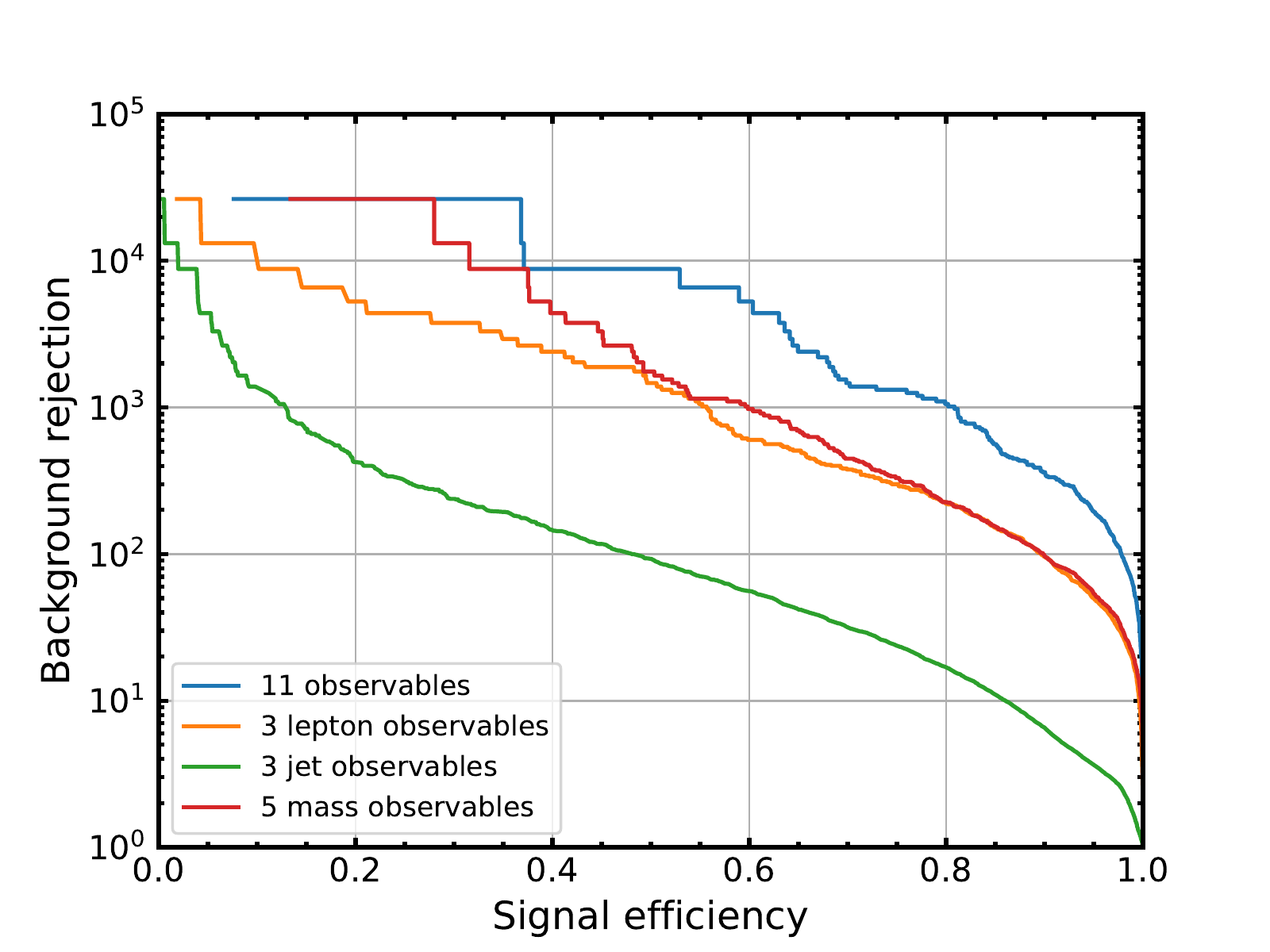}
\includegraphics[width=0.49\textwidth]{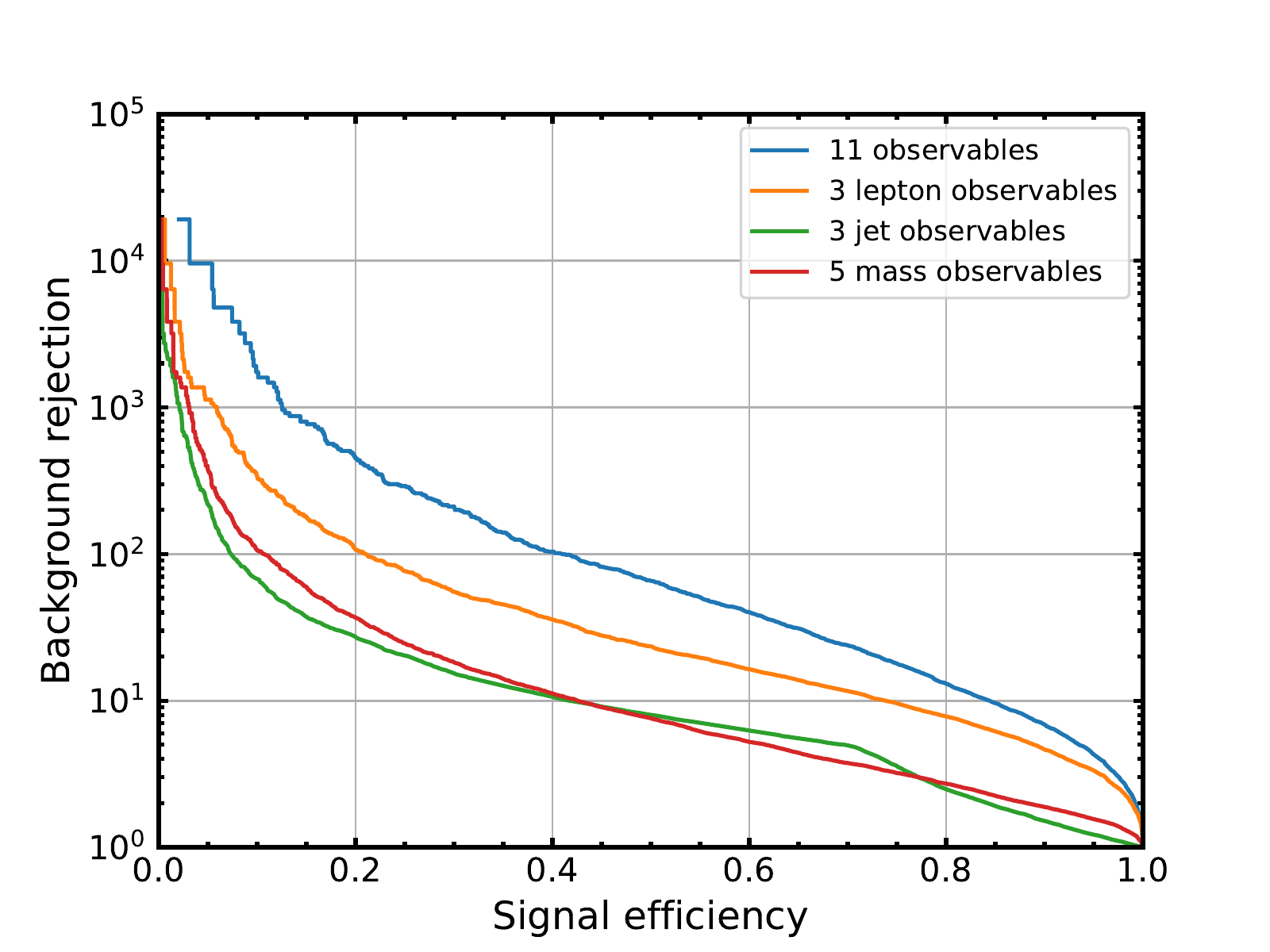}
\includegraphics[width=0.49\textwidth]{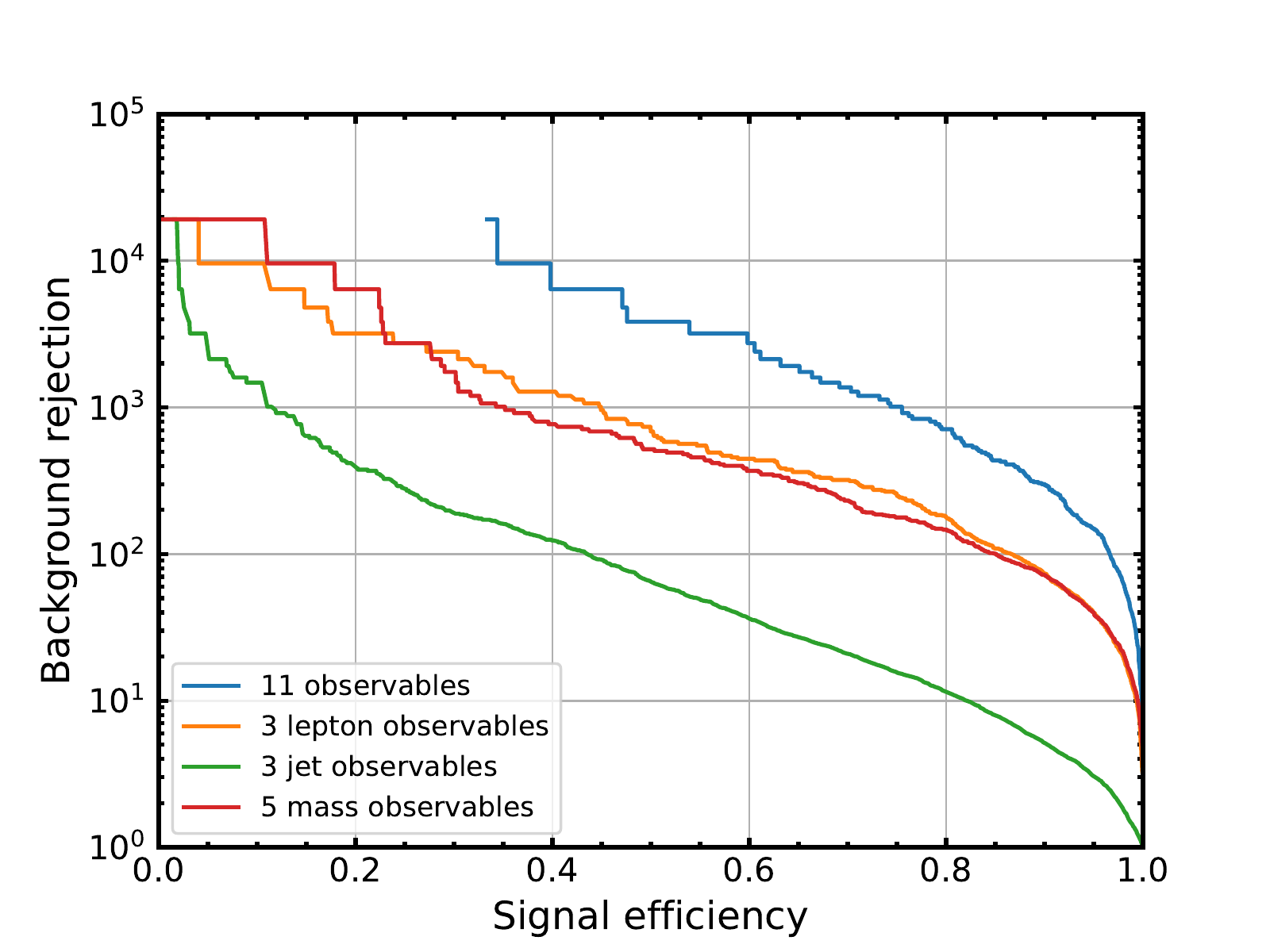}
\caption{Background rejection versus signal efficiency at the 13~(up), 27~(middle), 100~(bottom) TeV hadron colliders, with only 3 lepton, 3 jet, 5 mass observables, 
for mN = 200~(left), 800~(right) GeV.}
\label{fig:ML_bs_ob}
\end{figure}

Since we have 11 observables, it is interesting to compare them, to see how they contribute to identify the background and signal. Hence, In Fig.~\ref{fig:ML_bs_ob}, we show the background rejection versus signal efficiency at the 13~(up), 27~(middle), 100~(bottom) TeV hadron colliders, with only 3 lepton, 3 jet, 5 mass observables, 
for $m_N$ = 200~(left), 800~(right) GeV. The curves closer to the upright corner are better to distinguish the background from signal, since we have larger signal efficiency as well as background rejection in the same time. Overall, changing the collision energy leads to merely no differences. The all three categories of the observables have shown sound potential to distinguish the background from the signal. 
Comparing them, surprisingly, we find that the jet ones are the worst. This can be understood since the jets in the final states have experienced hadronization, shower and etc, therefore their kinematic is hard to reflect the different processes.
The contribution of the lepton and mass observables depends on the masses of the heavy neutrinos. 
For lighter heavy neutrinos, e.g. $m_N=$ 200 GeV as we show, lepton observables are the best. For heavier $N$, e.g. $m_N=$ 800 GeV, things become opposite, as mass observables now rule. It is because the mass observables are highly correlated to the masses of the heavy neutrinos. Thus heavier heavy neutrinos lead to significantly difference in them for signal and background.

\section{Sensitivity}
\label{sec:sen}

\begin{figure}[htbp]
\centering
\includegraphics[width=0.95\textwidth]{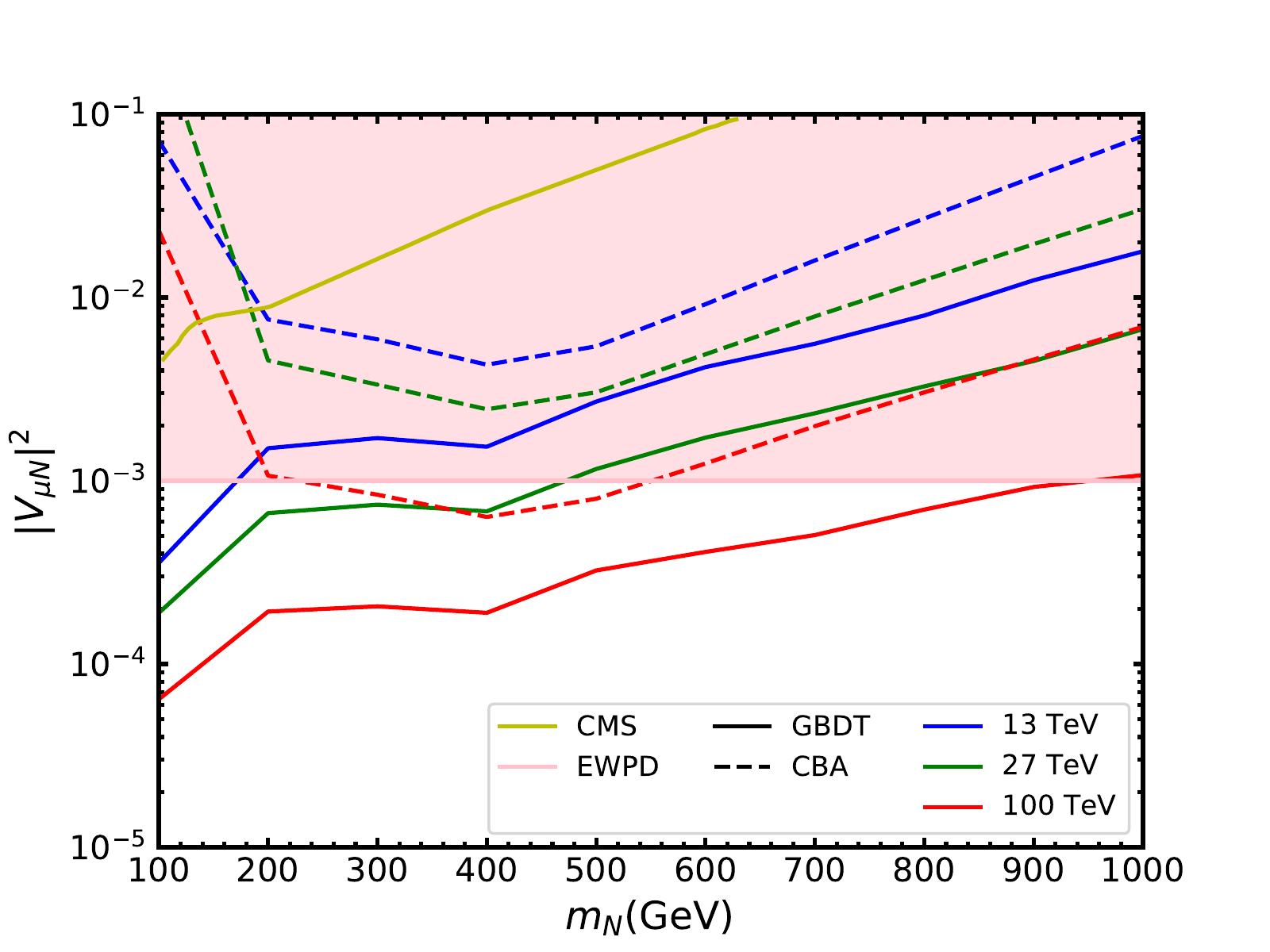}
\caption{The sensitivity reach on $(m_N, V_{\mu N}^2)$ plane at 95\% C.L. from the cut-based and machine learning analyses~(only GBDT) for the fat-jet signal of the heavy neutrinos, at the 13 TeV LHC with 3 ab$^{-1}$~(blue), 27 TeV TeV HE-LHC with 15 ab$^{-1}$~(green), and 100 TeV collider with 30 ab$^{-1}$ luminosities~(red). The existing limits from EWPD~\cite{delAguila:2008pw,Akhmedov:2013hec,Antusch:2014woa,Blennow:2016jkn} and CMS~\cite{CMS:2018iaf} are also shown for comparison.}
\label{fig:sen}
\end{figure}

After optimizing the signal efficiency and background rejection to get the best significance, we 
obtain the number of events for the signal and background processes, and derive the sensitivity for the relevant model parameters of the heavy neutrinos, $(m_N, V_{\mu N}^2)$. We use a Poisson distribution to set the sensitivity at 95\% confidence level~(C.L.), by requiring $S^2/(S+B) = 3.84 $~\cite{ParticleDataGroup:2020ssz}. 
The sensitivity reach on $(m_N, V_{\mu N}^2)$ plane at 95\% C.L. from the cut-based and machine learning analyses~(only GBDT) for the fat-jet signal of the heavy neutrinos, at the 13 TeV LHC with 3 ab$^{-1}$~(blue), 27 TeV TeV HE-LHC with 15 ab$^{-1}$~(green), and 100 TeV collider with 30 ab$^{-1}$ luminosities~(red) is shown in Fig.~\ref{fig:sen}. For ML analyses, we only show the results for the GBDT, since it can yield slightly better sensitivity comparing to the MLP, as Fig.~\ref{fig:ML_bs} suggests.
The sensitivity of the cut-based analyses at 13 TeV LHC roughly meets with the results in Ref.~\cite{Bhardwaj:2018lma}. Limits from other existing experiments are also overlaid for comparison, including the EWPD limits from the fact that the existence of heavy neutrinos can induce the non-unitary of the light neutrino mixing matrix~\cite{delAguila:2008pw,Akhmedov:2013hec,Antusch:2014woa,Blennow:2016jkn}, as well as the CMS searches for the lepton number conserving signals of the heavy neutrinos~\cite{CMS:2018iaf}.

With the help of ML techniques, we can probe $V_{\mu N}^2 \approx 5 \times 10^{-4}$ at the 13 TeV LHC when $m_N =$ 100 GeV, the sensitivity becomes worse since the signal cross section drops down for heavier $N$ with masses approaching the collision energy, reach to $V_{\mu N}^2 \approx 2 \times 10^{-2}$ when $m_N =$ 1 TeV. With higher collision energy, so the number of $W$ boson produced, the 27 TeV HE-LHC and 100 TeV future hardon colliders can lead us to where $V_{\mu N}^2 \approx 2 \times 10^{-4}$, $6 \times 10^{-5}$, respectively, at $m_N =$ 100 GeV. For $m_N =$ 1 TeV, we can still probe  $V_{\mu N}^2 \approx 6 \times 10^{-3}$ at the 27 TeV HE-LHC, and $V_{\mu N}^2 \approx 1 \times 10^{-3}$ at 100 TeV future hadron colliders.

The results of the cut-based analyses depend on how the threshold is chosen for each cuts. For example, the cuts on the invariant masses, $M(\mu^\pm J)>$ 500 GeV seems to be optimized for $m_N \approx$ 500 GeV, and we have obtain the best sensitivity around there. For the parameter far away, the threshold should be tuned. The ML analyses can overcome this difficulty, since only the distribution of the observables are required to fed to them, so the sensitivity curves are roughly smooth.

In Fig.~\ref{fig:sen}, it is clear that by employing the ML techniques, we can reach about one magnitude better sensitivity for the active-sterile mixing $V_{\mu N}^2$ as a function of the $m_N$, comparing to the cut-based analyses, and two magnitude better than the ones from the existing CMS searches. But the sensitivity at 13 TeV LHC seems to be worse than the indirect limits of the EWPD with $V_{\mu N}^2 \lesssim 10^{-3}$, only when $m_N <$~200 GeV. This can be overcome by taking higher collision energy, e.g. 27 and 100 TeV.
With the help of ML techniques, the sensitivity is comparable to the EWPD at the 13 TeV LHC, better when $m_N \lesssim$~500 GeV at the 27 TeV HE-LHC, and the whole mass range we concern at the 100 TeV hadron colliders.

Comparing the results to the ones in Ref.~\cite{Feng:2021eke}, we found that the sensitivity from fat-jet final states are slightly worse than the trilepton final states as they focused on. The fat-jet final states turn out to yield better sensitivity only when the heavy neutrinos are very heavy, $m_N \sim $ 1 TeV. In other parameter space, it seems that the fat-jet final states is worse than the trilepton for about 0.2-1 magnitude, especially when $m_N \lesssim$ 200 GeV. Although the hadronic final states have larger decay branching ratio of $N$, this advantage can not cancel out the large SM background of them. The situation is already optimized by extracting jet substructure information to the ML analyses as done in this paper, but still can not compete with the trilepton final states. At high-energy lepton colliders, i.e. the international linear collider~(ILC)~\cite{ILC:2013jhg}, the Compact Linear Collider (CLIC)~\cite{CLIC:2016zwp} and multi-TeV muon collider~\cite{Delahaye:2019omf}, the SM background for the fat-jet signal is much smaller, e.g. $\bar{t}t$ processes. We expect that the ML analyses for the fat-jet signal can yield better sensitivity to the trilepton final states. As already shown in Ref.~\cite{Chakraborty:2018khw} at CLIC.

\section{Conclusion}
\label{sec:con}
The search for the explanation of the observed neutrino masses stands as the one of the most important challenges at the current era of the particle physics. The heavy neutrinos in the inverse seesaw framework, can be an answer to that question. They can be Dirac, Majorana fermions, or the mixture of them. 
Towards this problem, we focus on the Dirac heavy neutrinos in this paper, and use jet substructure information to distinguish the signal from the background, by applying cut-based analyses, as well as ML analyses including the MLP and GBDT strategy for the $pp \rightarrow W^{\pm *}\rightarrow \mu^{\pm} N \rightarrow \mu^{\pm} \mu^{\mp} W^{\pm} \rightarrow \mu^{\pm} \mu^{\mp} J$ process. Comparing the two strategy, the GBDT has shown better potential in improving the signal over background ratio.

The observables of the final states including the kinematics of the leptons, fat-jet as well as the (transverse)~invariant masses of variable system are used. The cut-based analyses have been performed by setting threshold on each of them, while the ML using multi-dimensional analyses. We further compare the different observables of the ML analyses, and finding out that the observables for the fat-jet do hold powerful ability to separate signal from the background. The (transverse)~invariant masses of variable system, are proved to be the most significant one when $N$ is heavy. When the $N$ is light, the lepton ones are the most competent.

We have determined the sensitivity at the 13 TeV LHC, 27 TeV HE-LHC and 100 TeV future hadron colliders towards this signal processes, using ML analyses. The reach on the parameter space, have arrived at about $V_{\mu N}^2 \lesssim 10^{-3.3~(-3.7,-4.2)}-10^{-1.7~(-2.2,-3.0)}$ at the 13~(27,100) TeV hadron colliders for 100 GeV$< m_N<$ 1 TeV. This is about one magnitude better than the ones obtained by cut-based analyses, showing the strong ability of the ML techniques in finding the new physics. The ML methods is especially useful, when $N$ is light, as shown that their sensitivity is almost 2 magnitude better than the ones for the cut-based. Since it is hard to distinguish the background and signal from invariant masses there, while still can be rescued by using the multi-dimensional analyses of the ML methods.
Equipped with the ML techniques, 13 TeV LHC can shown better sensitivity comparing to the EWPD when $m_N \lesssim$ 200 GeV, and the 27 and 100 TeV hadronic colliders can exceed the EWPD in a much larger parameter space. Due to the large background of the hadronic final states, the sensitivity is  worse than the ones from the trilepton final states. We expect hadronic final states at High energy lepton colliders, especially the multi-TeV muon colliders can yield better sensitivity to the trilepton final states, which we leave for future studies.

\begin{acknowledgments}
WL is supported by National Natural Science Foundation of China (Grant No.12205153), and the 2021 Jiangsu Shuangchuang (Mass Innovation and Entrepreneurship) Talent Program (JSSCBS20210213). HS is supported by the National Natural Science
Foundation of China (Grant No.12075043, No.12147205).
\end{acknowledgments}

\bibliographystyle{JHEP}
\bibliography{submit.bib}
\end{document}